# A Local and Discrete Model

# Simulating Nonrelativistic Quantum Mechanical Systems

Antonio Sciarretta

34 rue du Château, Rueil Malmaison, France

Tel.: +39.147525734, E-mail: asciatopo@gmail.com

## Abstract

*This paper presents a simple model that mimics quantum mechanics (QM) results without using complex wavefunctions or non-localities. The proposed model only uses integer-valued quantities and arithmetic operations, in particular assuming a discrete spacetime under the form of a Euclidean lattice. The proposed approach describes individual particle trajectories as random walks. Transition probabilities are simple functions of a few quantities that are either randomly associated to the particles during their preparation, or stored in the lattice nodes they visit during the walk. Non-relativistic QM predictions are retrieved as probability distributions of similarly-prepared ensembles of particles. The scenarios considered to assess the model comprise of free particle, constant external force, harmonic oscillator, particle in a box, and the Delta potential.*

## 1 Introduction

Non-classical features of quantum mechanics like self-interference and Born's rule are described in terms of abstract mathematical objects. Although some have been ready to interpret the complex-valued wavefunction as a real object, wavefunctions are generally seen as mathematical tools serving to calculate probabilities from their square moduli. Contrasting to real-valued mathematics and one-to-one mapping between real variables and observables of classical theories, the standard description is thus sometimes considered as purely operational.

This paper investigates the possibility to develop a model that predicts probability fields of nonrelativistic quantum mechanical systems avoiding both complex-valued quantities and non-localities.

Actually, the model found shall even avoid real-valued quantities and use only integer-valued primary quantities (in addition to rational-valued quantities derived from them) to be combined using only arithmetic operations. As a matter of fact, a discrete spacetime under the form of a lattice shall be used. Individual particle trajectories shall be described as asymmetric random walks, with transition probabilities being simple functions of a few quantities that are either randomly associated to the particles during their preparation, or stored in the lattice nodes they visit during the walk. Non-relativistic QM predictions shall be retrieved as probability distributions of similarly-prepared ensembles of particles.

The idea to simulate quantum mechanics with random walks has its origin from the path integral formalism and the chessboard model introduced by R. Feynman [1]. Since then, many papers have appeared in the literature to retrieve the Schrödinger equation from a random walk process, including work of G.N. Ord [2-4] and subsequent refinements [5], as well as different yet related approaches [6]. While these approaches are able to reproduce the emergence of Schrödinger equation, the Born probability rule and interference are not explained in such models.

Following E. Nelson's seminal work [7], another approach has consisted of showing the emergence of the Schrödinger equation from a stochastic equation of motion [8-13]. Such derivation is founded on the assumption of reversible diffusion or competing diffusion-antidiffusion processes, leading to a key osmotic velocity that is non-local in the sense that it depends on the probability field it concurs in building, thus playing a similar role than quantum potential in De Broglie–Bohm mechanics and related theories [14, 15].

In contrast to the work above, the proposed model is aimed at predicting both Schrödinger equation and non-classical consequences of Born rule (as double-slit interference) only from the random walk features and the local interactions between the particles and the lattice.

Unlike other corpuscular models [16] the proposed model is not restricted to double-slit interference as it is not explicitly based on ad-hoc trigonometric functions inspired by the complex waveforms structure. Additionally, the proposed model uses the lattice only as the support for particle motion, not for wavefunctions or other mathematical operators. Finally, interference is not reproduced by appealing to probability cancellation due to antiparticles, nor to other definitions of negative probabilities.

The paper is organized as follows. First, the one-dimensional lattice is presented (Sect. 2) alongside with the fundamental spacetime quantization. Emission of particle at sources (Sect. 3) and particle motion (Sect. 4) are subsequently described. Then Schrödinger equation is retrieved by analysing the probability density functions of ensembles of particle emissions (Sect. 5). Finally, numerical simulations allow a comparison between the proposed model and quantum mechanical results for several scenarios (Sect. 6).

## 2 Lattice

The proposed model assumes a discrete spacetime. For simplicity, the description below will be limited to one dimension. The spatial values are thus restricted to integer multiples of a fundamental quantity $X$ and the temporal values are restricted to integer multiples of a quantity $T$. In the rest of the paper, except when explicitly stated, these integer values will be denoted with small Latin letters, while the corresponding physically-valued quantities will be generally denoted with a tilde.

In this model, a particle's evolution consists of the succession of discrete values $x[n]$, $t[n]$, where $n \in \mathbb{N}$ is the index that describes advance in history, here denoted as "iteration". By taking an arbitrary $x = 0$ reference, the spacetime may be thought as if it is constituted by a grid $x \in \mathbb{Q}, t \in \mathbb{N}$, or "lattice", whose nodes can be visited by the particle during its evolution.

Advance in time is unidirectional and unitary, that is,

$$t[n] = t[n-1] + 1\,, \qquad t[n_0] = 0\,, \tag{1}$$

where $n_0$ is the iteration when the particle is created. Advance in space is still unitary but the particle can either advance in one of the two directions or stay at rest, according to the rule

$$x[n+1] = x[n] + v[n] , \qquad x[n_0] = x_0 , \tag{2}$$

where the motion is regulated by a random variable $v$ ("momentum") that can take only three values, namely, $v \in \{-1,0,1\}$.

The fundamental quantities $X$ and $T$ are related to the Compton length and time,

$$X = \frac{h}{2mc} , \qquad T = \frac{h}{2mc^2} , \tag{3}$$

where $m$ is the particle's mass, $c$ is the speed of light, and $h$ is Planck constant. Relations (3) are the same as those introduced by G.N. Ord and the authors of [2-6]. Here, however, a slightly different justification is proposed as follows.

From the rule (2), it is clear that the particle reaches its maximum speed when $v[n] \equiv 1$, which provides, in physical units,

$$\frac{X}{T} = c . \tag{4}$$

On the other hand, consider the random variable defined as the particle's sample momentum after $N$ iterations of observation, $v^{(N)}$. For $n = 1$, the possible outcomes of $\tilde{v}^{(1)}$ are (in physical units) $-c, 0, +c$. Thus the uncertainty $\Delta\tilde{v}^{(1)}$ is equal to $c$. For $N = 2$, the possible outcomes are $-c, -c/2, 0, c/2, c$, with $\Delta\tilde{v}^{(2)} = c/2$. After $N$ observations, $\Delta\tilde{v}^{(N)} = c/N$. However, observing the particle for $N$ iterations implies an uncertainty in the determination of its position as well. Since the position might change from $-NX$ to $NX$, $\Delta\tilde{x}^{(N)}$ is equal to $2NX$ (in physical units). Multiplying the uncertainty in the particle's momentum ($mv$) and that in the particle's position, obtain

$$m\Delta\tilde{v}^{(N)}\Delta\tilde{x}^{(N)} = 2mcX = h , \tag{5}$$

which is compliant with Heisenberg uncertainty principle. Actually, the latter would imply a factor $\hbar$ at the right-hand side of (5); however, in a one-dimensional space the factor $2\pi$ (half the solid angle of a sphere) is correctly replaced in the proposed model by the factor 1 (half the measure of the unit 1-sphere). From (4) and (5), (3) results.

## 3 Particle emissions

Particles are generally created (or "emitted") at a source node ("source") $x_0$, which can be fixed or randomly determined according to a pmf that represents real scenarios. Two pieces of information are attributed to a particle during its emission: a "source momentum" and a "source phase".

The former, $v_0 \in \mathbb{Q}$, is a uniformly-distributed rational-valued random variable ($v_0 \approx U[-1,1]$) that is determined at the preparation and does not change during the particle's evolution. This source momentum plays a key role in the proposed model in introducing an intrinsic randomness into the particle's evolution.

The source phase, $\epsilon \in \mathbb{Q}$, is a property of the source node and does not change during particle's evolution. The particular function $\epsilon(x_0)$ is set in such a way to represent real scenarios. In most cases, it shall give place to a drift momentum that is summed to $v_0$.

# 4 Microscopic motion

In this section the general characteristics of the random variable $v$ introduced in (2) are described. Since $v$ can take only three values at each time step, its probability distribution is completely determined by two values, its expected value and its variance. In the rest of the paper, expected values will be denoted in bold.

Define the "momentum propensity" as $\boldsymbol{v} \coloneqq E[v]$. The model further assumes that

$$E[v^2] \coloneqq \frac{1+\boldsymbol{v}^2}{2} := \boldsymbol{e}\,, \tag{6}$$

or, in other terms, $\mathrm{Var}[v] = (1-\boldsymbol{v}^2)/2$. Both $\boldsymbol{v}$ and $\boldsymbol{e}$ are not integers but rational numbers (this point will be clarified later). It should be noticed that, since $\boldsymbol{v} \in [-1,1]$, also $\boldsymbol{e} \in [-1,1]$. The symbol $\boldsymbol{e}$ recalls the fact that this quantity can be regarded as the average value of instantaneous particle's energy and will be denoted as "energy propensity".

Consequently to (6), the probability distribution of $v$ is determined as

$$\Pr(v=1) = \frac{\boldsymbol{e}+\boldsymbol{v}}{2}\,, \qquad \Pr(v=0) = 1-\boldsymbol{e}\,, \qquad \Pr(v=-1) = \frac{\boldsymbol{e}-\boldsymbol{v}}{2}\,. \tag{7}$$

The quantity $\boldsymbol{v}$ is the result of two mechanisms: (i) imprint during the particle's "preparation" at the source before its emission, and (ii) iteration-by-iteration evolution according to two types of forces, namely, "quantum forces" and "external forces". In summary,

$$\boldsymbol{v}[n] := v_Q[n] + v_F[n]\,, \qquad \boldsymbol{v}[n_0] = v_0\,, \tag{8}$$

where $v_Q$ is the contribution due to quantum forces, $v_F$ is the contribution due to external forces. It should be noticed that, according to (8) and further rules below, $\boldsymbol{v}$ is a random rational-valued variable as anticipated.

In the absence of either quantum or external forces, the proposed model is summarized as

$$M_0 \coloneqq \begin{cases} t[n+1] = t[n] + 1\,, \qquad t[n_0] = 0\,, \\ x[n+1] = x[n] + v[n]\,, \qquad x[n_0] = x_0\,, \\ \Pr(v = 0, \pm 1) = (7)\,, \\ \boldsymbol{v}[n] = v_0\,, \\ v_0 = U[-1;1]\,. \end{cases} \tag{9}$$

## 4.1 External forces

External forces are described by interactions with the lattice, where each node can be occupied by momentum-mediating entities that will be called "bosons" in analogy with physical force-mediating particles. Depending on their origin, these bosons have an intrinsic momentum propensity $v_f$. The probability of finding such a boson at a certain node, $P_f(x,t)$, depends on the rate at which such bosons are emitted by their source and the distance from the source (the time dependency is because boson source can be variable).

This fundamental mechanism is equivalent to, and for computational easiness replaced by, the following one: bosons are always available at each node where $P_f \neq 0$ and have an intrinsic momentum propensity $f(x,t) \coloneqq v_f P_f(x,t)$. The particle passing by the lattice node captures the resident boson and incorporates its momentum. A new boson is then recreated at the lattice node.

The contribution to the particle's momentum propensity due to external forces is thus given by the sum of the momenta of all external bosons captured,

$$v_F[n] = \sum_{n'=n_0+1}^{n} f(x[n'], n') \,. \tag{10}$$

It should be noticed that equation (10) is analogous to classical Newton's law in lattice units.

Under the sole action of external forces, the proposed model is summarized as

$$M_1 := \begin{cases} t[n+1] = t[n] + 1 \,, \qquad t[n_0] = 0 \,, \\ x[n+1] = x[n] + v[n] \,, \qquad x[n_0] = x_0 \,, \\ \Pr(v = 0, \pm 1) = \ (7) \,, \\ \boldsymbol{v}[n+1] = \boldsymbol{v}[n] + f(x[n], n) \,, \qquad \boldsymbol{v}[n_0] = v_0 \,, \\ v_0 = U[-1;1] \,. \end{cases} \tag{11}$$

## 4.2 Quantum forces

The contribution to the momentum propensity due to quantum forces is given by

$$v_Q[n] = v_0 - \sum_i \sum_{j \neq i} v_Q^{(ij)}[n] \,, \tag{12}$$

where each term in the summation at the right-hand side of (12) results from an exchange of information between the particle and the lattice. In fact, both the particle and the lattice nodes carry and store some integer-valued "counters" that can be updated as iterations proceed.

The counters carried on by the particle are its lifetime, $t[n]$, and a spatial counter $\ell[n]$ denoted as "span". The counter stored at each lattice node $\xi$ is the "trace" $\lambda_{\xi\tau}[n]$ of the span carried by the last particle that has visited the node with lifetime $\tau$. The dynamics of these counters are given by

$$\ell[n] = \begin{cases} \lambda_{x[n]t[n]}[n-1], & \text{if } \ell[n] \neq \lambda_{x[n]t[n]}[n-1] \\ \ell[n-1] + v[n], & \text{else if } f(x[n], n) = 0 \\ -\ell[n-1], & \text{otherwise} \end{cases} \,, \qquad \ell[n_0] = 0 \,, \tag{13}$$

$$\lambda_{\xi\tau}[n] = \begin{cases} \ell[n], & \text{if } \mathsf{C}_{\xi\tau} \\ \lambda_{\xi\tau}[n-1], & \text{otherwise} \end{cases} . \tag{14}$$

where the reset condition for the lattice counter is $\mathsf{C}_{\xi\tau} := (x[n] = \xi) \wedge (t[n] = \tau) \wedge \big(\ell[n] \neq \lambda_{\xi\tau}[n-1]\big)$. As for the particle span $\ell$, it is generally updated at each iteration by summing to its previous value the value of the instantaneous momentum. However, if the particle finds a lattice node with a trace that is different from its span, the two counters are interchanged. Additionally, if the particle experiences an external force, the span has its sign reversed.

According to these rules, it should be clear that the trace found by a particle can be different from its span because the last particle that visited the node with the same lifetime either had been emitted from a different source $x_0$ or had captured a different number of external bosons. In any case, it should be noticed that $\ell[n] \in \mathbb{Z}$. Consequently, also $\lambda_{\xi\tau}[n] \in \mathbb{Z}$.

The interchange between the particle span and the lattice trace is accompanied by the creation of a new momentum-carrying "lattice boson". This boson is labelled with the particular couple of integers $\ell[n]$, $\lambda_{\xi\tau}[n]$ or, equivalently, with the couple $ij$, where $i = \xi - \ell[n]$ and $j = \xi - \lambda_{\xi\tau}[n-1]$. Clearly, $i \in \mathbb{Z}$ and $j \in \mathbb{Z}$ are images of the respective sources of the current particle and of the last particle that has visited the node $\xi$.

The lattice boson is created with a momentum that equals the quantum momentum of the visiting particle and replaces the previously resident boson of the same type, if there was one. The momentum of the latter, before to be replaced, is however passed to the particle as the contribution $v_Q^{(ij)}$ in the right-hand side of (11). In other words, the lattice and the particle boson of the same type exchange their momenta. Additionally, both momenta decay with the respective bosons' lifetimes: at a new iteration, the momentum value is only a fraction of the previous value.

The mechanism can be formalized as follows. The particle's boson dynamics is given by

$$v_Q^{(ij)}[n] = \begin{cases} v_Q^{(ij)}[n-1] \cdot \left(1 - \dfrac{1}{2k^{(ij)}[n]}\right), & \text{if } k^{(ij)}[n] > 0 \\ \omega_{x[n]t[n]}^{(ij)}[n], & \text{otherwise} \end{cases}, \tag{15}$$

$$k^{(ij)}[n] = \begin{cases} 0, & \text{if } \mathrm{C}^{(ij)} \\ k^{(ij)}[n-1] + 1, & \text{else} \end{cases}, \tag{16}$$

where $k^{(ij)}$ is the lifetime of the *ij*-boson and the reset condition for $k^{(ij)}$ (a new particle boson to be created) is $\mathrm{C}^{(ij)} := (\ell[n] = x[n] - i) \wedge \left(\lambda_{x[n]t[n]}[n-1] = x[n] - j\right)$.

The quantity $\omega_{\xi\tau}^{(ij)}[n]$ is the momentum carried by the lattice *ij*-boson. Its dynamics is given by the rules

$$\omega_{\xi\tau}^{(ij)}[n] = \begin{cases} \omega_{\xi\tau}^{(ij)}[n-1] \cdot \left(1 - \left(\dfrac{\delta^{(ij)} \varpi_{\xi\tau}^{(ij)}[n]}{\kappa_{\xi\tau}^{(ij)}[n]}\right)^2\right), & \kappa_{\xi\tau}^{(ij)}[n] > 0 \\ \varpi_{\xi\tau}^{(ij)}[n], & \text{otherwise} \end{cases}, \tag{17}$$

$$\varpi_{\xi\tau}^{(ij)}[n] = \begin{cases} \varpi_{\xi\tau}^{(ij)}[n-1], & \kappa_{\xi\tau}^{(ij)}[n] > 0 \\ v_Q[n] - \dfrac{\epsilon^{(ij)}}{\delta^{(ij)}}, & \text{otherwise} \end{cases}, \tag{18}$$

$$\kappa_{\xi\tau}^{(ij)}[n] == \begin{cases} 0, & \text{if } \mathrm{C}_{\xi\tau}^{(ij)} \\ \kappa_{\xi\tau}^{(ij)}[n-1] + 1, & \text{otherwise} \end{cases}, \tag{19}$$

where $\kappa_{\xi\tau}^{(ij)}$ is the lifetime of the lattice *ij*-boson, $\varpi_{\xi\tau}^{(ij)}$ is its initial momentum, and the reset condition for $\kappa_{\xi\tau}^{(ij)}$ (condition for a new lattice boson to be created) is $\mathrm{C}_{\xi\tau}^{(ij)} := (x[n] = \xi) \wedge (t[n] = \tau) \wedge (\ell[n] = \xi - i) \wedge \left(\lambda_{\xi\tau}[n-1] = \xi - j\right)$.

The additional quantities introduced above are the path difference

$$\delta^{(ij)} := |i - j| = \left|\ell[n] - \lambda_{\xi\tau}[n]\right| \tag{20}$$

and the phase difference

$$\epsilon^{(ij)} := \epsilon(i) - \epsilon(j)\,, \tag{21}$$

so that $\epsilon^{(ij)}/\delta^{(ij)}$ describes a "phase momentum" resulting from a different preparation at the two sources. It should be noticed that the rules above preserve the fact that $v_Q \in \mathbb{Q}$, $\varpi_{\xi\tau}^{(ij)} \in \mathbb{Q}$, and $v_Q^{(ij)} \in \mathbb{Q}$.

The complete set of equations of the proposed model is summarized as follows:

$$M \coloneqq \begin{cases} t[n] = t[n-1] + 1\,, \qquad t[n_0] = 0\,, \\ x[n+1] = x[n] + v[n]\,, \qquad x[n_0] = x_0\,, \\ \Pr(v = 0, \pm 1) = \ (7)\,, \\ \boldsymbol{v}[n] \coloneqq v_Q[n] + v_F[n]\,, \qquad \boldsymbol{v}[n_0] = v_0\,, \\ v_0 = U[-1;1]\,, \\ v_F[n] = \displaystyle\sum_{n'=n_0+1}^{n} f(x[n'], n')\,, \\ v_Q[n] = v_0 - \displaystyle\sum_i \sum_{j \neq i} v_Q^{(ij)}[n]\,, \\ v_Q^{(ij)}[n] = \ (15) - (19)\,. \end{cases} \tag{22}$$

## 5 Probability densities

In this section some general features of the random variables introduced above are calculated. Even without quantum or external interactions, the fact that the source momentum is a random variable implies that $\boldsymbol{v}$ and thus $\boldsymbol{x}$ are random variables, too. The probability mass function $\rho(\boldsymbol{x}; t)$ is calculated from the probability mass functions of the quantum momentum $\rho(v_Q)$ and that of the source momentum, i.e.,

$$\rho(v_0) = \frac{1}{2}\,. \tag{23}$$

Additionally, the source location is treated as a random variable, too. In general, there are $N_s$ possible sources, located at nodes $x_0^{(k)}$, each of which has a probability $P_0^{(k)}$. In other terms,

$$x_0 = \left\{x_0^{(k)}\right\}\,, \qquad k = 1, \dots, N_s\,, \qquad \rho(x_0) = P_0^{(k)} \cdot \delta\left(x - x_0^{(k)}\right), \tag{24}$$

where $\delta(\cdot)$ is Dirac delta function. Five special cases are considered for the sake of presentation: (i) no forces, (ii) only quantum forces, (iii) only homogeneous external forces from a quadratic potential, (iv) quantum and homogeneous external forces from a quadratic potential, (v) quantum and inhomogeneous external forces. It turns out that for all these cases the expected value of the position is a monotonic function of the quantum momentum and the latter of the source momentum. The chain rule

$$\rho(\boldsymbol{x}; t) = \rho(v_0)\left|\frac{dv_0}{d\boldsymbol{x}}\right| = \rho(v_0)\left|\frac{dv_0}{dv_Q}\right|\left|\frac{dv_Q}{d\boldsymbol{x}}\right| \tag{25}$$

is then applied.

## 5.1 No forces

The only scenario without forces acting on the particle is when there is a single source possible and no external forces. In this scenario, each lattice node $\xi$ always receives particles carrying a span equal to $\xi - x_0$, so that no bosons are created.

For illustration purposes, the probability mass function can be explicitly calculated in this case. For a given $\boldsymbol{v} = v_0$, the pmf of $x$ at a given $t$ is

$$\rho_v(x;t) = \frac{\binom{2t}{t+x-x_0}}{2^{2t}}(1+\boldsymbol{v})^{t+x}(1-\boldsymbol{v})^{t-x} \tag{26}$$

and is well approximated by a Gaussian function $\frac{1}{\sqrt{2\pi\mathcal{D}t}}\exp\left(-\frac{(x-x_0-vt)^2}{2\mathcal{D}t}\right)$, where $\mathcal{D} := (1-\boldsymbol{v}^2)/2$. By integrating over values of $v_0$, obtain

$$\rho(x;t) = \int_{-1}^{1}\rho(v_0)\rho_v(x;t)dv_0 = \frac{1}{2t+1}, \tag{27}$$

that is, a constant pmf in the reachable interval, $x = U[x_0 - t, x_0 + t]$.

The same result can be approximated by using (25) and observing that $\boldsymbol{x} = x_0 + v_0 t$ in this case, therefore

$$\rho(\boldsymbol{x};t) = \frac{1}{2t}. \tag{28}$$

It should be noticed that $\rho(x;t) \approx \rho(\boldsymbol{x};t)$ for large times.

For such a simple scenario it is also possible to explicitly compute the pmf of another random variable defined for each lattice node as

$$\sigma_{\xi\tau}[n] := \begin{cases} s[n], & \text{if } (x[n] = \xi) \wedge (t[n] = \tau) \\ \sigma_{\xi\tau}[n-1], & \text{otherwise} \end{cases}, \tag{29}$$

The particle random variable $s$ is in turn defined recursively as

$$s[n+1] = s[n] + |v[n]|, \qquad s[n_0] = 0. \tag{30}$$

The variable $s$ can be regarded as the accumulated energy of the particle, in agreement with the fact that the expected value of $|v|$ is the energy propensity $\boldsymbol{e}$ defined above, and will be referred to here as the particle's "action", at least for this special case (a term due to external bosons is actually missing). The variable $\sigma_{\xi\tau}$ is the particle action "seen" by the node when particles visit it. It should be further noticed that both $s \in \mathbb{N}$ and $\sigma_{\xi\tau} \in \mathbb{N}$.

The pmf of the latter can be explicitly calculated for this simple scenario as

$$\rho_v(\sigma;x,t) = \frac{2^{t-s}\binom{t}{\frac{\sigma+x-x_0}{2}}\binom{t-\frac{\sigma+x-x_0}{2}}{t-\sigma}}{\binom{2t}{t+x-x_0}}, \tag{31}$$

where $\sigma \in \left\{|x-x_0|, |x-x_0|+2, \ldots, |x-x_0| + 2\left\lfloor\frac{t-|x-x_0|}{2}\right\rfloor\right\}$ (subscripts $\xi\tau$ have been omitted here for the sake of clarity). Since the right-hand side of (31) does not depend on $v_0$, $\rho(\sigma;x,t) = \rho_v(\sigma;x,t)$ holds as well.

The expected value of the action seen at a node is found with some algebraic manipulation to be

$$\boldsymbol{\sigma}_{xt} = \frac{(x-x_0)^2 + t^2 - t}{2t-1}, \tag{32}$$

which, for large times, is remarkably similar to the classical free particle action $S(x,t) = \frac{(x-x_0)^2}{2t}$ plus the term $t/2$.

In conclusion, the proposed model approximates for large times the probability density and the action of a free particle emitted from a single source, albeit only using integer and rational quantities.

## 5.2 Quantum forces only

When source location can take multiple values, quantum forces occur. In fact, a lattice node $\xi$ can receive particles carrying a span that takes either of the values $\xi - x_0^{(k)}$, so that bosons are created.

Consider the generic $ij$-bosons. They are created at a node when $t[n] = \tau$, $\ell[n] = \xi - x_0^{(i)}$, and $\lambda_{\xi\tau}[n-1] = \xi - x_0^{(j)}$. As far as $\omega_{\xi\tau}^{(ij)}$ is concerned, it should be noticed that it depends on the boson's lifetime according to the decay rule (17). By repeatedly applying such a rule for $\kappa$ iterations, obtain

$$\omega_{\xi\tau}^{(ij)}(\kappa) = \bar{\omega}_{\xi\tau}^{(ij)} \prod_{\kappa'=1}^{\kappa} \left(1 - \left(\frac{\delta^{(ij)}\bar{\omega}_{\xi\tau}^{(ij)}}{\kappa'}\right)^2\right). \tag{33}$$

The steady-state value of the lattice boson momentum is obtained by letting $\kappa$ tend to infinity and is a function of $\bar{\omega}_{\xi\tau}^{(ij)}$, i.e., $\sin\left(\pi\delta^{(ij)}\bar{\omega}_{\xi\tau}^{(ij)}\right)/\pi\delta^{(ij)}$, using the known formula for the sine expansion, $\sin\pi z = \pi z \prod_{n=1}^{\infty}\left(1 - \frac{z^2}{n^2}\right)$. It should be remarked that a trigonometric functionality emerges from the integer-valued model proposed.

The initial lattice boson momentum $\bar{\omega}_{\xi\tau}^{(ij)}$ is determined according to rule (18). It should be clear that its expected value is thus

$$\bar{\boldsymbol{\omega}}_{\xi\tau}^{(ij)} = \frac{\left(\xi - x_0^{(i)}\right) + \left(\xi - x_0^{(j)}\right)}{2\tau} - \frac{\epsilon^{(ij)}}{\delta^{(ij)}} = \frac{\xi - \dfrac{x_0^{(i)} + x_0^{(j)}}{2}}{\tau} - \frac{\epsilon^{(ij)}}{\delta^{(ij)}} = \bar{\boldsymbol{\omega}}_{\xi\tau}^{(ji)}. \tag{34}$$

Consequently,

$$\boldsymbol{\omega}_{\xi\tau}^{(ij)} = \frac{\sin\left(\pi\delta^{(ij)}\bar{\boldsymbol{\omega}}_{\xi\tau}^{(ij)}\right)}{\pi\delta^{(ij)}}. \tag{35}$$

Now consider the particle $ij$-boson. Its momentum varies with its lifetime, according to rule (15). Now, the probability that such a boson has lifetime $k$ is equal to the probability that in $k$ iterations a new boson is created only once. The probability that a new boson is created equals that of the joint event $P^{(ij)} \coloneqq P_0^{(i)}P_0^{(j)} = P^{(ji)}$. Therefore, $\Pr\left(k^{(ij)} = k\right) = P^{(ij)}\left(1 - P^{(ij)}\right)^k$. The expected value of $v_Q^{(ij)}$ is now calculated as

$$\boldsymbol{v}_{\boldsymbol{Q}}^{(\boldsymbol{ij})} = P^{(ij)} \sum_{k=0}^{\infty} v_Q^{(ij)}(k) \cdot \left(1 - P^{(ij)}\right)^k , \tag{36}$$

where $v_Q^{(ij)}(k)$ denotes now the boson momentum with lifetime $k$. By repeatedly applying rule (15), obtain

$$v_Q^{(ij)}(k) = v_Q^{(ij)}(0) \cdot \prod_{k'=1}^{k} \frac{2k'-1}{2k'} . \tag{37}$$

The product in (37) is calculated as $\frac{(2k)!}{(k!)^2 4^k}$ that, for the properties of Gamma function, is formally equivalent to $(-1)^k \binom{-1/2}{k}$. Therefore, (36) is manipulated as

$$\begin{aligned} \boldsymbol{v}_{\boldsymbol{Q}}^{(\boldsymbol{ij})} &= P^{(ij)} \sum_{k=0}^{\infty} \left(1 - P^{(ij)}\right)^k \cdot v_Q^{(ij)}(0) \cdot (-1)^k \binom{-1/2}{k} = \\ &= v_Q^{(ij)}(0) P^{(ij)} \left(1 - \left(1 - P^{(ij)}\right)\right)^{-1/2} = v_Q^{(ij)}(0) \sqrt{P^{(ij)}} , \end{aligned} \tag{38}$$

using the binomial series expansion $(1-z)^\alpha = \sum_{k=0}^{\infty} \binom{\alpha}{k} (-z)^k$. The final step consists of replacing $v_Q^{(ij)}(0)$ with $\boldsymbol{\omega}_{\boldsymbol{xt}}^{(\boldsymbol{ij})}$, according to rule (15) and with the change of subscripts $\xi \to \boldsymbol{x}$ to the lattice boson momentum. Using (35), find

$$\boldsymbol{v}_{\boldsymbol{Q}}^{(\boldsymbol{ij})} = \sqrt{P^{(ij)}} \frac{\sin\left(\pi \delta^{(ij)} \overline{\boldsymbol{\omega}}_{\boldsymbol{xt}}^{(\boldsymbol{ij})}\right)}{\pi \delta^{(ij)}} . \tag{39}$$

The expected value of the particle's quantum momentum is eventually found as

$$\boldsymbol{v}_{\boldsymbol{Q}} = v_0 - \sum_i \sum_{j \neq i} \sqrt{P_0^{(i)} P_0^{(j)}} \frac{\sin\left(\pi \delta^{(ij)} \frac{\boldsymbol{x} - \frac{x_0^{(i)} + x_0^{(j)}}{2}}{t} - \pi \epsilon^{(ij)}\right)}{\pi \delta^{(ij)}} . \tag{40}$$

The pmf of the position cannot be explicitly calculated in this case. However, the probability density function of its expected value can be calculated, using (40) and observing that $\boldsymbol{x} = x_0 + \boldsymbol{v}_{\boldsymbol{Q}} t$ holds in this case, as

$$\begin{aligned} \rho(\boldsymbol{x}; t) &= \frac{1}{2} \frac{dv_0}{d\boldsymbol{x}} = \\ &= \frac{1 + \sum_i \sum_{j \neq i} \sqrt{P_0^{(i)} P_0^{(j)}} \cos\left(\pi \delta^{(ij)} \frac{\boldsymbol{x} - \frac{x_0^{(i)} + x_0^{(j)}}{2}}{t} - \pi \epsilon^{(ij)}\right)}{2t} . \end{aligned} \tag{41}$$

A further analysis of (41) remarkably allows to retrieve Schrödinger equation. First replace $\delta^{(ij)}$ with $\left| x_0^{(i)} - x_0^{(j)} \right|$ and then recognize that the cosine argument is equal to

$$\frac{2\boldsymbol{x}\left(x_0^{(i)} - x_0^{(j)}\right) - \left(\left(x_0^{(i)}\right)^2 - \left(x_0^{(j)}\right)^2\right)}{2t} - \epsilon^{(ij)} = -\Delta S^{(ij)}(\boldsymbol{x}, t) . \tag{42}$$

where $S^{(k)}(x,t)$ is the classical action with respect to the $k$-th source,

$$S^{(k)}(x,t) \coloneqq \frac{\left(x - x_0^{(k)}\right)^2}{2t} + \epsilon^{(k)} . \tag{43}$$

The right-hand side of (41) can be then equivalently obtained as the square modulus of a complex number, $\rho(\boldsymbol{x};t) = |\psi(x,t)|^2$, where

$$\psi(x,t) = \sum_k \psi^{(k)}(x,t), \qquad \psi^{(k)}(x,t) \coloneqq \sqrt{\frac{P_0^{(k)}}{2t}} \exp\left(\iota\pi S^{(k)}(x,t)\right) . \tag{44}$$

It is easy to recognize in $\psi(x,t)$ the probability amplitude of the free particle for many possible sources, each of which has a probability amplitude $\psi^{(k)}(x,t)$. With this observation, the equivalence between the proposed model and quantum mechanics is demonstrated for the scenario considered.

## 5.3 Homogeneous external forces only (quadratic potentials)

This section treats the scenario where the initial distribution of particles is again as they are emitted from a single source; however, particles are now subject to external forces. The analysis is limited to quadratic potentials such that the external boson momentum is

$$f(x,t) = \alpha(t)x + \beta(t) . \tag{45}$$

It should be noticed that (45) imposes no restrictions on $x$. Consequently, each node visited by the particle transmits an external boson and, according to rule (14),

$$\ell[n] = (-1)^{\mathrm{parity}(t[n])} \cdot \frac{(x[n] - x_0)}{t[n]} . \tag{46}$$

The span carried by the particle at a node only depends on its lifetime and thus there are no possible differences between the span and the trace found that might be induced by external forces. Consequently, quantum forces are always null, $v_Q[n] \equiv v_0$ and $v[n] = v_0 + v_F[n]$.

Computing the pdf of $\boldsymbol{x}$ requires the particularization of the function $f(x,t)$ that describes the external boson momentum. Generally speaking, for quadratic potentials it is always possible to write

$$\boldsymbol{x}(t) = A(t)x_0 + B(t)v_0 + C(t) , \tag{47}$$

where $A(t)$, $B(t)$, and $C(t)$ are functions of lifetime whose form depends on the coefficients $\alpha$ and $\beta$ of (45), such as

$$\begin{cases} A(0) = 1 , & \dot{A}(0) = 0 , \quad \ddot{A}(t) = \alpha(t)A(t) , \\ B(0) = 0 , & \dot{B}(0) = 1 , \quad \ddot{B}(t) = \alpha(t)B(t) , \\ C(0) = 0 , & \dot{C}(0) = 0 , \quad \ddot{C}(t) = \alpha(t)C(t) + \beta(t) , \end{cases} \tag{48}$$

and

$$\dot{A}B = A\dot{B} - 1 . \tag{49}$$

Some examples of external forces will be presented in Sect. 4. In general terms, the pdf of the particle position can be calculated as

$$\rho(\boldsymbol{x};t) = \frac{1}{2}\frac{dv_0}{d\boldsymbol{x}} = \frac{1}{2B(t)} . \tag{50}$$

The accessible domain of the particle position is limited by the trajectories obtained by setting $v_0 = \pm 1$ in (47). Consequently, it can be verified that

$$\int_{A(t)x_0+B(t)+C(t)}^{A(t)x_0-B(t)+C(t)} \rho(\boldsymbol{x};t)d\boldsymbol{x} = 1\,. \tag{51}$$

## 5.4 Quantum and homogeneous external forces (quadratic potentials)

In this scenario the particle encounters both quantum and external bosons. Equation (47) is replaced by

$$\boldsymbol{x}(t) = A(t)x_0 + B(t)\boldsymbol{v_Q} + C(t)\,, \tag{52}$$

where $x_0$ is, as in Sect. 3.2, a random variable.

The application of rule (18) yields

$$\overline{\boldsymbol{\omega}}_{\boldsymbol{xt}}^{(\boldsymbol{ij})} = \frac{v_Q^{(i)} + v_Q^{(j)}}{2} - \frac{\epsilon^{(ij)}}{\delta^{(ij)}} = \frac{\boldsymbol{x} - A(t)\frac{x_0^{(i)} + x_0^{(j)}}{2} - C(t)}{B(t)} - \frac{\epsilon^{(ij)}}{\delta^{(ij)}}, \tag{53}$$

and consequently, using the same reasoning as in Sect. 3.2, find

$$\boldsymbol{v_Q} = \\ = v_0 - \sum_i \sum_{j\neq i} \sqrt{P_0^{(i)} P_0^{(j)}} \frac{\sin\left(\pi\delta^{(ij)} \frac{\boldsymbol{x} - A(t)\frac{x_0^{(i)} + x_0^{(j)}}{2} - C(t)}{B(t)} - \pi\epsilon^{(ij)}\right)}{\pi\delta^{(ij)}}, \tag{54}$$

as the generalization of (40). The final step is to generalize (41) and find

$$\rho(\boldsymbol{x};t) = \\ = \frac{1 + \sum_i \sum_{j\neq i} \sqrt{P_0^{(i)} P_0^{(j)}} \cos\left(\pi\delta^{(ij)} \frac{\boldsymbol{x} - A(t)\frac{x_0^{(i)} + x_0^{(j)}}{2} - C(t)}{B(t)} - \pi\epsilon^{(ij)}\right)}{B(t)}. \tag{55}$$

Similarly to Sect. 3.2, it should be noticed that the cosine argument can be expressed as

$$\frac{2\left(x_0^{(i)} - x_0^{(j)}\right)\boldsymbol{x} - A(t)\left(\left(x_0^{(i)}\right)^2 - \left(x_0^{(j)}\right)^2\right) - 2C(t)\left(x_0^{(i)} - x_0^{(j)}\right)}{2B(t)} \\ - \epsilon^{(ij)} = -\Delta S_{cl}^{(ij)}(\boldsymbol{x},t)\,, \tag{56}$$

where

$$S_{cl}^{(k)}(x,t) = \frac{1}{2B(t)}\Big[A(t)\left(x_0^{(k)}\right)^2 - 2xx_0^{(k)} + 2C(t)x_0^{(k)} + \dot{B}(t)x^2 \\ + \left(2\dot{C}(t)B(t) - 2\dot{B}(t)C(t)\right)x + 2C^2(t)\dot{B}(t)\Big] + \epsilon^{(k)} \tag{57}$$

is the classical action in the presence of the considered potential, as it can be easily verified.

The right-hand side of (57) can be then equivalently obtained as the square modulus of a complex number, $\rho(\boldsymbol{x};t) = |\psi(x,t)|^2$, where

$$\psi(x,t)=\sum_k \psi^{(k)}(x,t), \qquad \psi^{(k)}(x,t) \coloneqq \sqrt{\frac{P_0^{(k)}}{2t}}\exp\left(\iota\pi S_{cl}^{(k)}(x,t)\right). \qquad (58)$$

It is easy to recognize in $\psi(x,t)$ the probability amplitude of the particle for many possible sources, each of which has a probability amplitude $\psi^{(k)}(x,t)$. With this observation, the equivalence between the proposed model and quantum mechanics is demonstrated also for the scenario considered.

### 5.5 Additional scenarios

It is worth mentioning the fact that equation (55) and its less general counterparts are valid also when the possible sources can be approximated as a continuum (Gaussian waves, stationary states, etc.). In these cases, sums can be conveniently approximated as integrals over distinct pairs of sources.

If the initial state presents a phase, as it is the case of, for instance, propagating Gaussian waves, a corresponding source phase $\epsilon$ must be taken into account in the model. In order to be consistent with quantum mechanics, the latter is taken as the angle of the QM initial amplitude divided by $\pi$.

The more complex scenario presented here is when the force field, i.e., the function $f(x,t)$, does not concern all the possible lattice nodes $x$. In such cases, the sign of the span $\ell$ depends on the path taken, and more precisely on the number of external bosons encountered. Therefore, even for particles emitted from a single source, different spans can be monitored at a given lattice node. As a consequence, quantum forces arise. Generally speaking, these problems are equivalent to introducing a certain number of image sources, each with its own probability. Special cases of this type will be treated in Sect. 4.

## 6 Numerical results

As a general feature of the simulations presented in this section, the model equations are repeated for a series of $N_p$ consecutive particle emissions, so that the motion of each particle is simulated for $N_t$ iterations. Probability density function of position is retrieved as the frequency of arrivals.

The stabilisation of quantum mechanisms and the emergence of quantum-like behaviour require a large number of emissions $N_p$ (and large times $N_t$). In order to speed up the calculations and make their reasonable for personal computers, it is assumed that (i) the lattice is already trained after a large number of previous, non-simulated emissions, and (ii) the particle is also trained.

Lattice training is the process during which $\omega_{\xi\tau}^{(ij)}$ tends to its expected value. This process is illustrated in Figure 2, where random variables $\overline{\omega}_{\xi\tau}^{(ij)}$ and $\omega_{\xi\tau}^{(ij)}$ are plotted versus the number of iterations for a few emissions. It is clear that $\omega_{\xi\tau}^{(ij)}$ has generally enough time between two successive emissions to converge to a steady-state value depending on $\overline{\omega}_{\xi\tau}^{(ij)}$. Now, $\overline{\omega}_{\xi\tau}^{(ij)}$ can change only at time $\tau$ of each emission. It is clear that after a sufficiently large number of iterations the sample mean of $\overline{\omega}_{\xi\tau}^{(ij)}$ tends to its expected value (34) and thus that of $\omega_{\xi\tau}^{(ij)}$ to (35).

Particle training is the process during equality (39) is build up. This process is illustrated in Figure 3, where one random variable $v_Q^{(ij)}$is shown versus the number of iterations for one emission, together with its running average and the quantity $\boldsymbol{v}_{\boldsymbol{Q}}^{\boldsymbol{ij}}$ expected from (39). The figure clearly shows that the running average tends after a sufficiently long time to the expected value. Consequently, also the "average momentum" calculated as $(x - x_0)/t$ tends to $\boldsymbol{v_Q}$ given by (40).

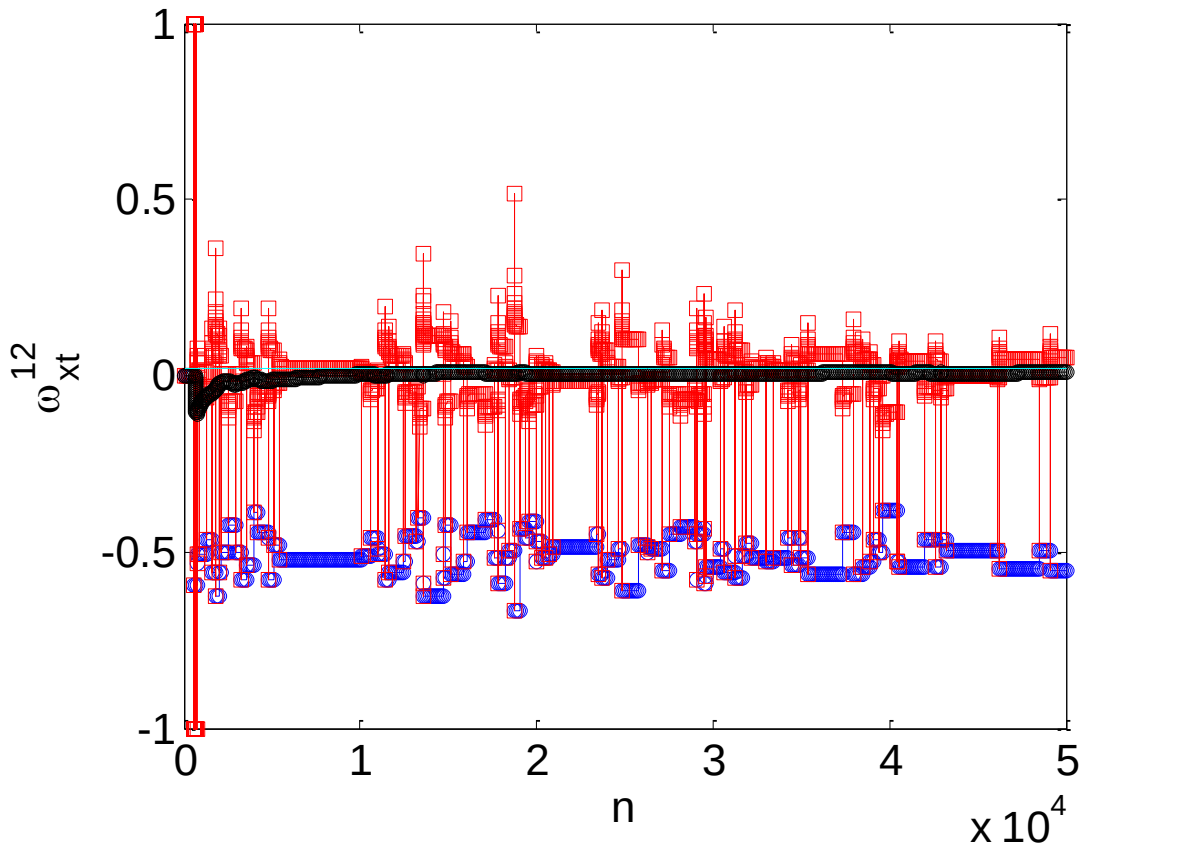

Figure 1: Outcome of one simulation ($x_0 = \{100 \pm 1\}$, $P_0 = \{0.5,0.5\}$) in terms of $\varpi_{xt}^{(12)}$ (blue), $\omega_{xt}^{(12)}$ (red), its running average (black), and $\boldsymbol{\omega_{xt}^{(12)}}$ (cyan) for a node ($x = 48$, $t = 100$) as a function of the number of iterations.

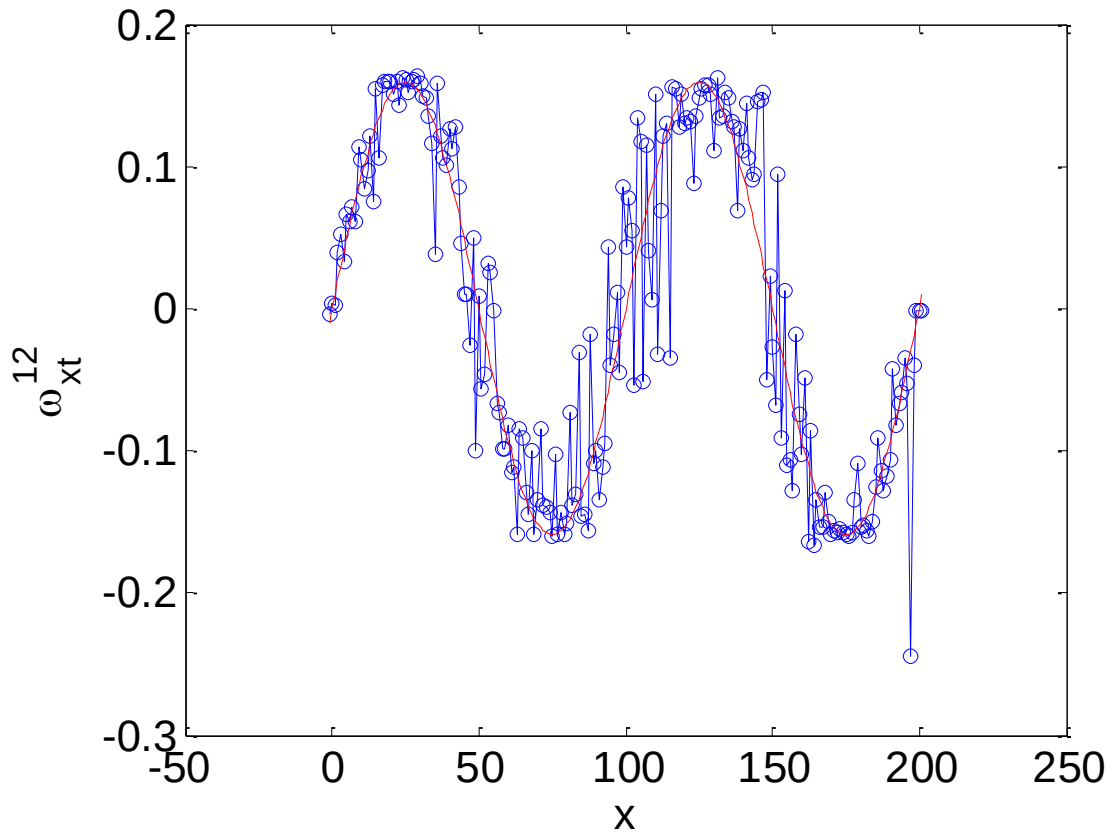

Figure 2: Outcome of the same simulation of Figure 1 (after $N_p = 50000$) in terms of $\omega_{xt}^{(12)}$ (blue) and $\boldsymbol{\omega_{xt}^{(12)}}$ (red) for each lattice node.

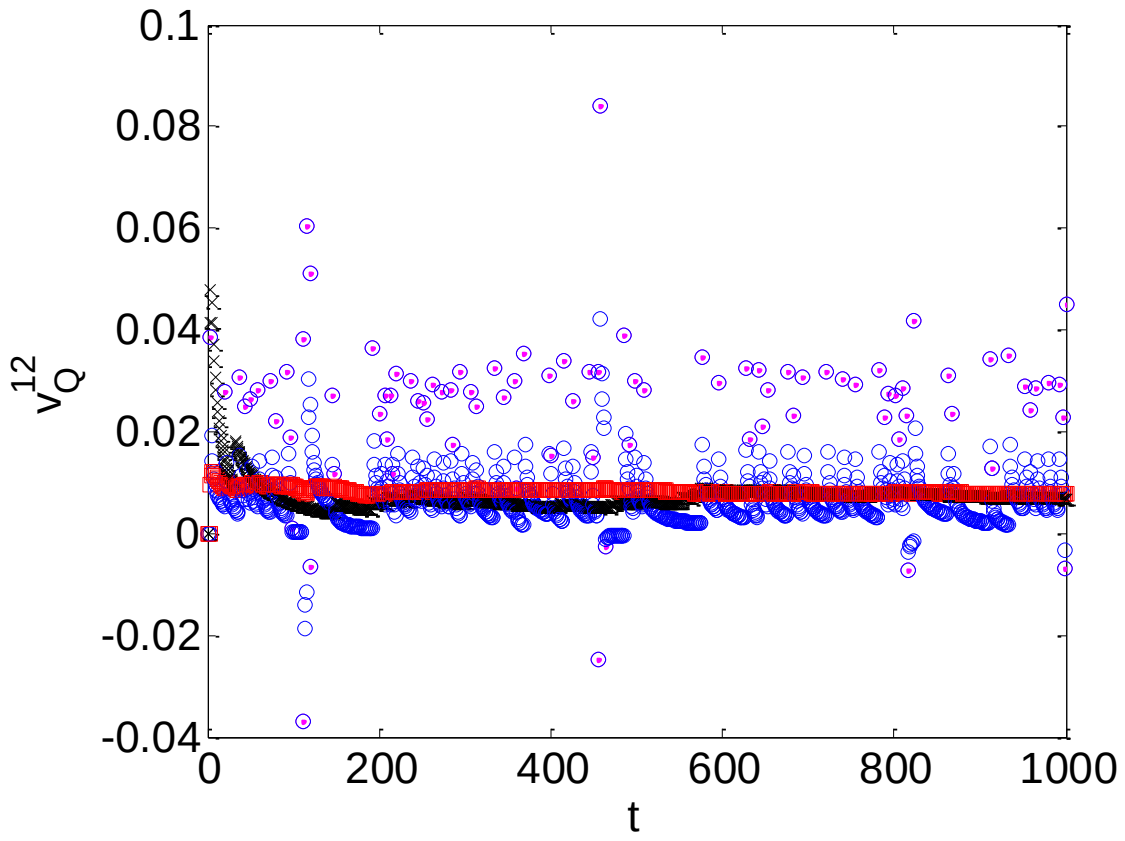

Figure 3: Outcome of one simulation ($N_t = 1000$, $v_0 = .7845$, $x_0 = \{\pm 1\}$, $P_0 = \{0.1,0.9\}$) in terms of $v_Q^{(12)}$ (blue), its running average (red), and $\boldsymbol{v_Q^{(12)}}$ (black) as a function of particle's lifetime.

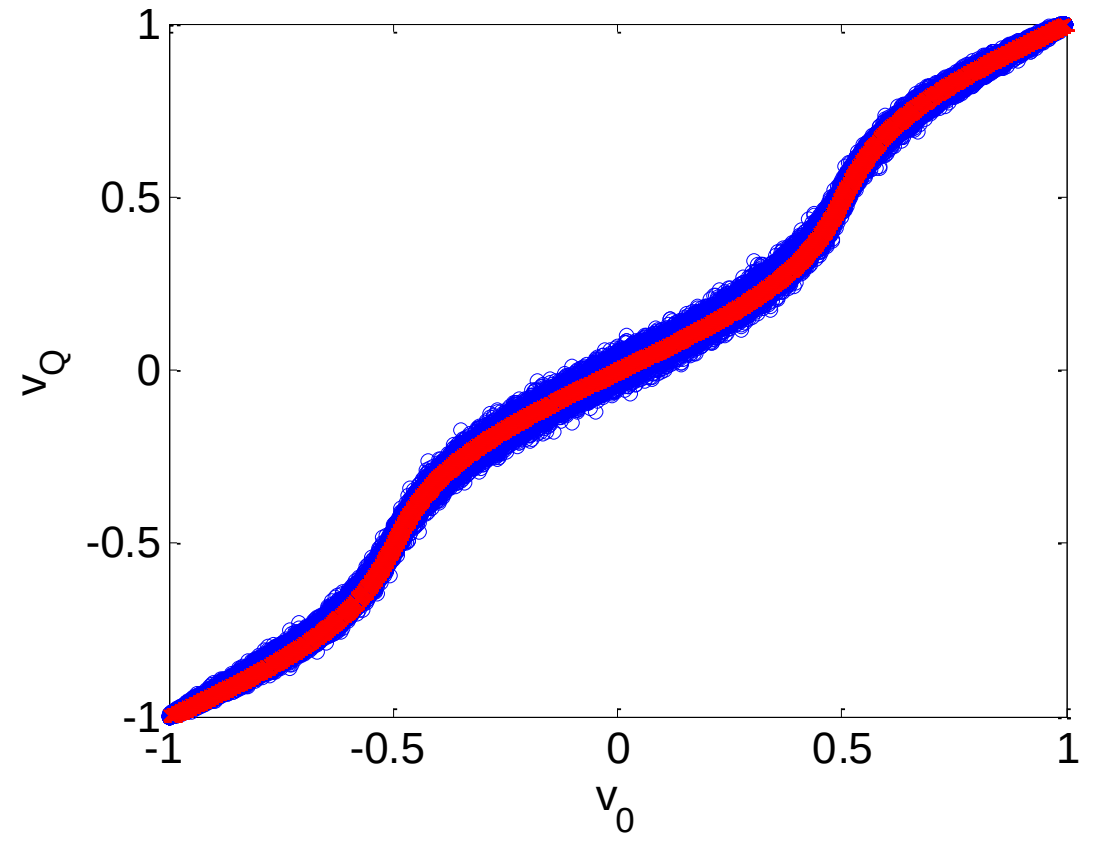

Figure 4: Outcome of the same simulation of Figure 3 (for $N_p = 50000$ emissions) in terms of "average momentum" $(x(N_t) - x0)/N_t$ (blue) and $\boldsymbol{v_Q}$ (red) as a function of $v_0$.

To dispose of an accelerated model that converges faster to quantum mechanical results, while taking into account the fact that lattice and particle "training" is not an instantaneous process, an artificial time lag of a few $n_\tau$ iterations is introduced such that the model (15)-(19) is replaced by its "trained" counterpart

$$v_Q^{(ij)}[n] = \frac{\sqrt{P^{(ij)}} \sin\left(\pi\left(\delta^{(ij)} q[n] - \epsilon^{(ij)}\right)\right)}{\pi \delta^{(ij)}}, \quad q[n] = q[n-1] \cdot \frac{(n_\tau - 1)}{n_\tau} + \frac{1}{n_\tau} \cdot v_Q[n]. \tag{59}$$

with the $P^{(ij)}$, $\delta^{(ij)}$, and $\epsilon^{(ij)}$ pre-computed for each $ij$ pairs.

For this paper, simulations have been carried on with both the lattice and the particles trained (with $n_\tau = t$). Results of simulations of several scenarios are presented in the Appendix A. Force scenarios comprise of free particle (17A.1), constant force (A.2), harmonic oscillator (A.3), particle in a box (A.4) and Delta potential (A.5). For each of these cases, various initial states are reproduced (single source, multiple equally distanced sources, Gaussian waves, stationary states, etc.). Results of the proposed model are compared with quantum mechanical predictions, computed by applying the respective amplitude propagators to the initial states selected.

# 7 Discussion

The example discussed in the Appendix A show the ability of the proposed model to reproduce quantum mechanical behaviour of ensembles of particles similarly prepared. One question that might arise if this ability is really due to the discrete nature of the model and the particle-lattice interaction proposed. In other terms, one might wonder if the quasi-deterministic, continuous-space model given by (1), (2), (8), (10), (12), and (59), with $v$ replaced by its expected value $\boldsymbol{v}$, would suffice. That would give a model $M^*$ summarized as

$$M^* := \begin{cases} t[n] = t[n-1] + 1\,, \qquad t[n_0] = 0\,, \\ x[n+1] = x[n] + \boldsymbol{v}[n]\,, \qquad x[n_0] = x_0\,, \\ \boldsymbol{v}[n] := v_Q[n] + v_F[n]\,, \qquad \boldsymbol{v}[n_0] = v_0\,, \\ v_F[n] = \displaystyle\sum_{n'=n_0+1}^{n} f(x[n'], n')\,, \\ v_Q[n] = v_0 - \displaystyle\sum_i \sum_{j \neq i} v_Q^{(ij)}[n]\,, \\ v_Q^{(ij)}[n] = \dfrac{\sqrt{P^{(ij)}} \sin\left(\pi\left(\delta^{(ij)} q[n] - \epsilon^{(ij)}\right)\right)}{\pi \delta^{(ij)}}\,, \\ q[n] = q[n-1] \cdot \dfrac{(n_\tau - 1)}{n_\tau} + \dfrac{1}{n_\tau} \cdot v_Q[n]\,. \end{cases} \tag{60}$$

with $P^{(ij)}$, $\epsilon^{(ij)}$, and $\delta^{(ij)}$ determined at the preparation together with $x_0$, and $v_0 = U[-1,1]$. Model $M^*$ would yield classical trajectories with additional forces and a random initial speed. On ensembles of particles, it would produce similar results than the model $M$ (22) in most cases.

However, it would not describe the penetration of particles in classically forbidden regions of space. For instance, scenario A.5 (Delta potential) and in particular the arrangement 0 (Gaussian wave) could not be properly simulated with this model: particles would tunnel through the potential barrier only for $v_Q^2$ large enough, contrarily to what expected.

Moreover, model $M^*$ (60) would be nonlocal, in the sense that a particle emitted at a certain source would know about other possible sources since the beginning of its evolution, which is in contrast with the requirements set in the Introduction.

These facts reveals that integer quantities and in particular discrete spacetime are necessary in $M$ for at least two reasons: (i) to establish a local exchange between particles and the lattice and establish quantum forces, and (ii) to introduce a certain probability for particles to tunnel through potential barriers regardless of their momentum. The arithmetic operations on the these integer quantities emerge mostly from probability rules or counter updates.

Several refinements of the model are still possible. On the one hand, relativistic Newton's second law shall inspire a mechanism to prevent that the momentum propensity becomes larger than unity under the action of persistent forces. Extension to two- and three-dimensional spaces is required too, although that seems rather natural to perform. A set of three momentum propensities, $\boldsymbol{v}_x$, $\boldsymbol{v}_y$, and $\boldsymbol{v}_z$ shall be introduced, fulfilling the condition that the total energy $\boldsymbol{e} = \left(1 + v_x^2 + v_y^2 + v_z^2\right)/2 \leq 1$. This condition implies that $\boldsymbol{v}_x^2 + \boldsymbol{v}_y^2 + \boldsymbol{v}_z^2 \leq 1$, thus fixing a constraint to the probability densities of the three source propensities.

# A. Appendix

## A.1 Free particle

For all the scenarios in this section, $f \equiv 0$. Results are compared with quantum mechanics (theoretical values) by using the propagator

$$K^{(FP)}(x,t|y) = \frac{1}{\sqrt{2\iota t}} \cdot \exp\left\{\frac{\iota\pi(x-y)^2}{2t}\right\}. \tag{61}$$

in lattice units.

**Single source**

In this case, $x_0 = \{x_0\}$, $P_0 = \{1\}$. In Sect. 3.1 the equivalence between the proposed model and quantum mechanics has been already demonstrated while obtaining (28), i.e.

$$\rho(\boldsymbol{x};t) = \frac{1}{2t} \Longrightarrow \boldsymbol{x} = U[x_0 - t, x_0 + t]. \tag{62}$$

Figure 5-Figure 6 show the frequency of arrivals after $N_t = 500$ iterations, obtained with an increasing number of emissions at $x_0 = 0$. As $N_p$ increases, the frequency clearly tends to the theoretical probability density (62).

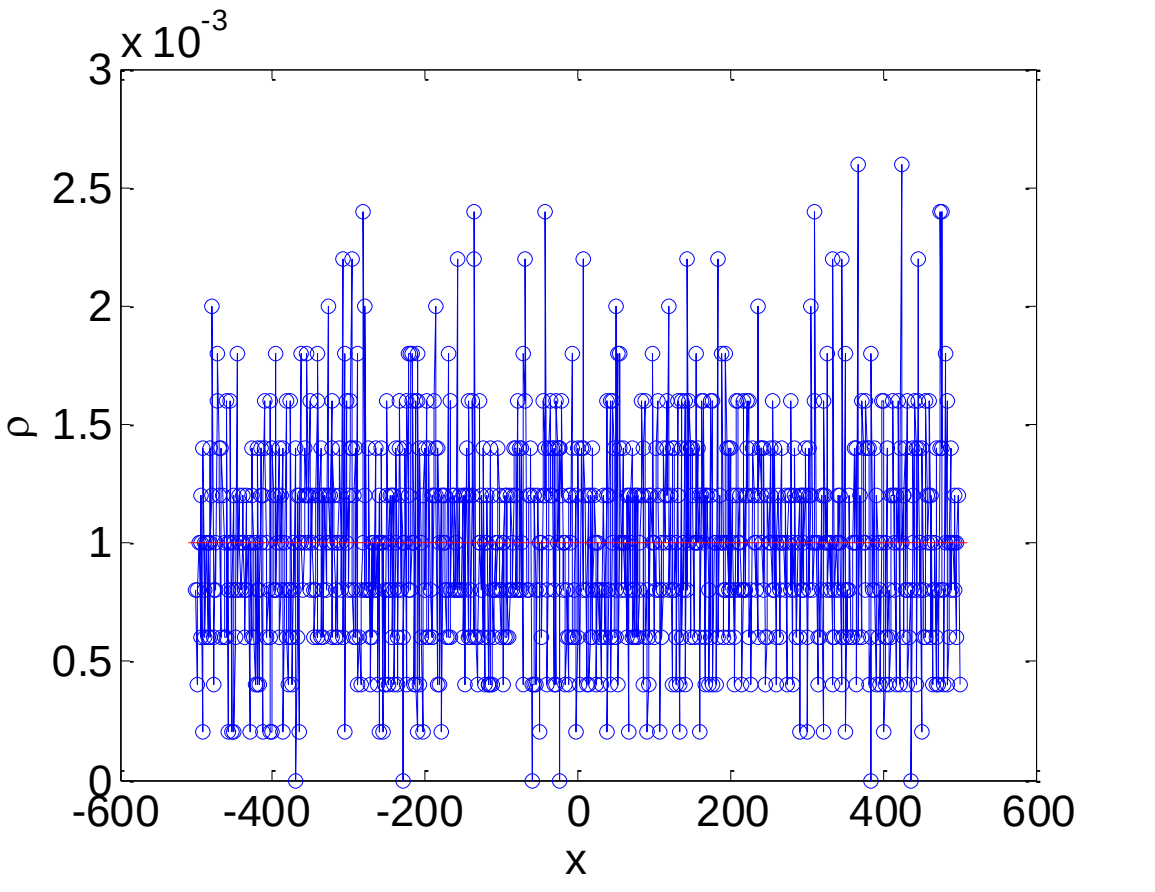

Figure 5: Arrival frequency (blue) and theoretical value (red) for $N_p = 5000$, $t = 500$ as a function of position (free particle, single source, $x_0 = 0$).

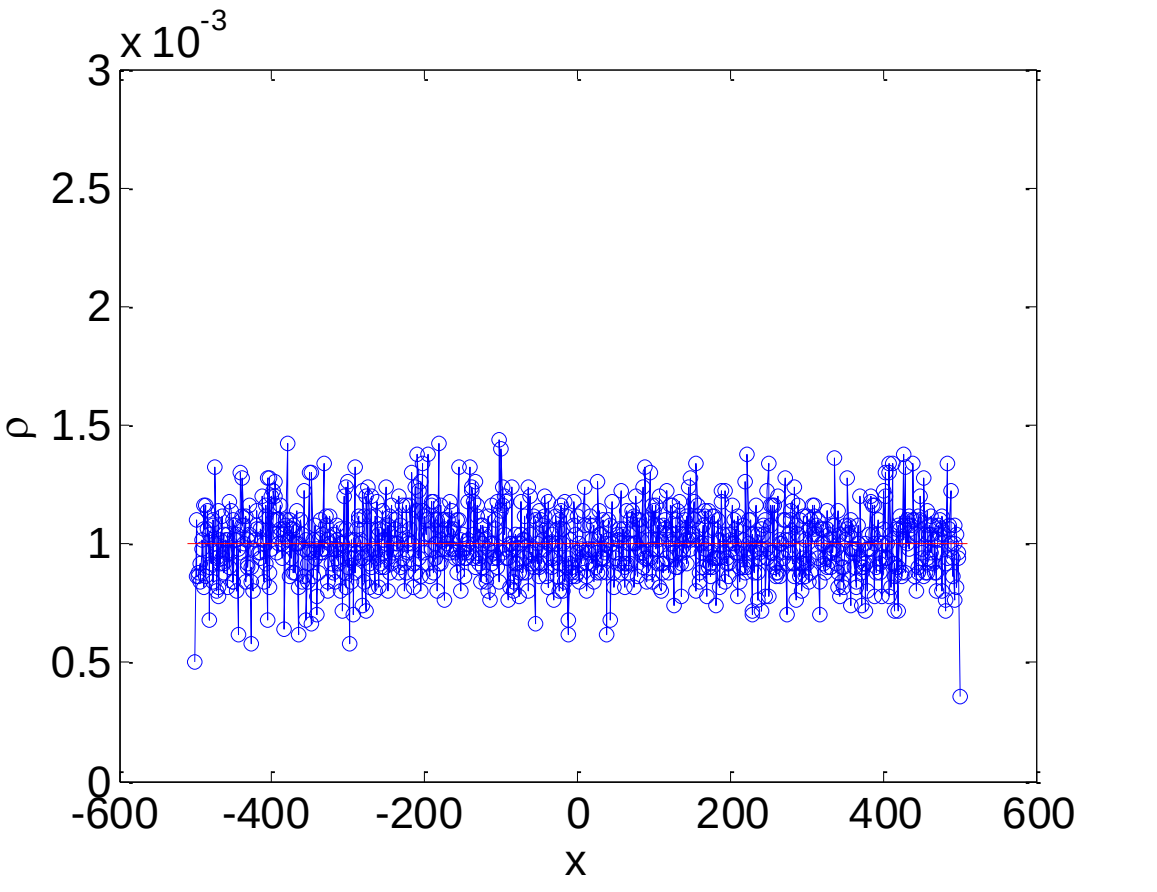

Figure 6: Arrival frequency (blue) and theoretical value (red) for $N_p = 50000$, $t = 500$ as a function of position (free particle, single source, $x_0 = 0$).

**Two sources**

This case is equivalent to the classical two-slit experiment, whereas $x_0 = \{-D, D\}$, $P_0 = \left\{P_0^{(-D)}, P_0^{(D)}\right\}$, with $P_0^{(D)} + P_0^{(-D)} = 1$. Consequently, two types of quantum boson arise $\delta^{(12)} = \delta^{(21)} = 2D$, while $\epsilon^{(12)} = \epsilon^{(21)} = 0$ (zero phase difference at the sources). The theoretically expected pdf is

$$\rho(\boldsymbol{x};t) = \frac{1 + 2\sqrt{P_0^{(D)}P_0^{(-D)}}\cos\left(2\pi D\frac{\boldsymbol{x}}{t}\right)}{2t}. \qquad (63)$$

Figure 9-Figure 10 show the frequency of arrivals after $N_t = 500$ iterations with emissions at $\pm D = \pm 1$ and with two different sets of source probabilities, $P_0^{(D)} = 1/2$ and $P_0^{(D)} = 0.1$, respectively. In both cases, the frequency clearly tends to the theoretical probability density (63).

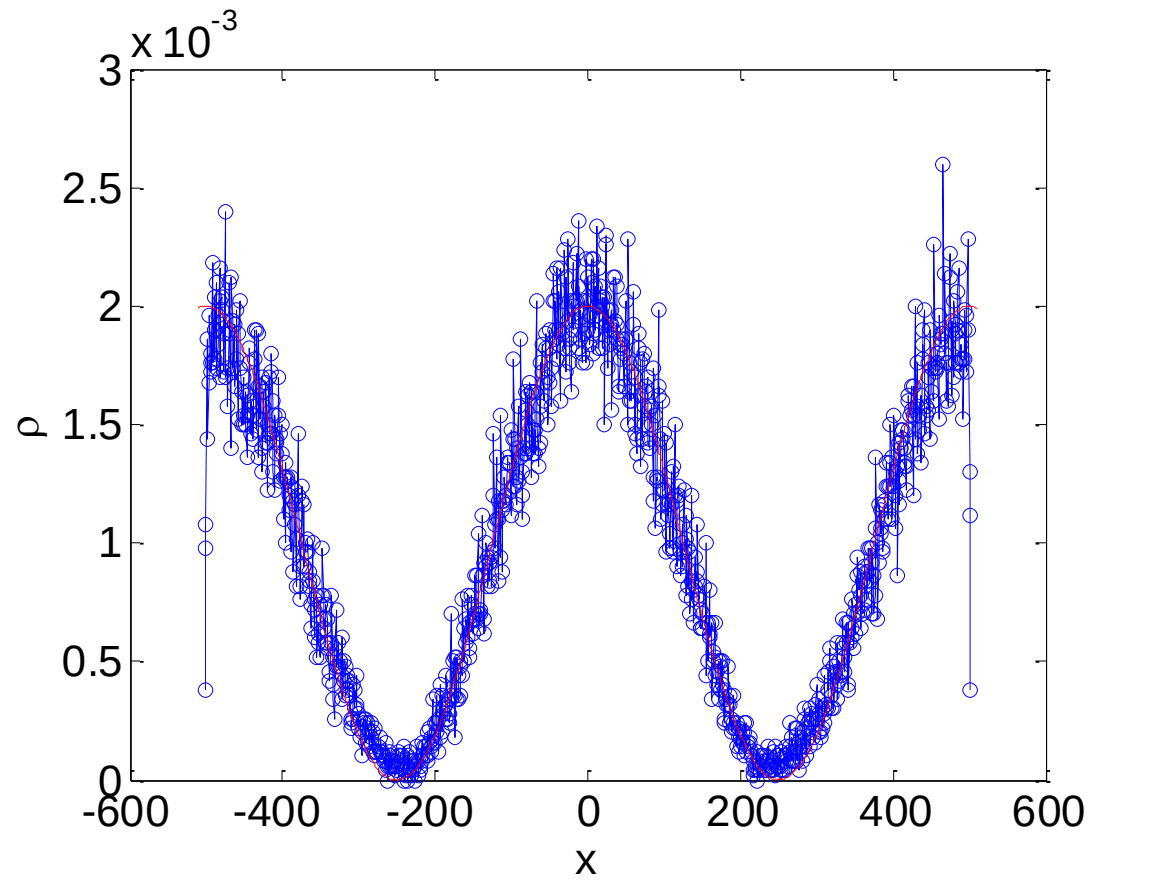

Figure 7: Arrival frequency (blue) and theoretical value (red) for $N_p = 50000$, $t = 500$ as a function of position (free particle, two sources, $D = 1$, $P_0^{(D)} = .5$).

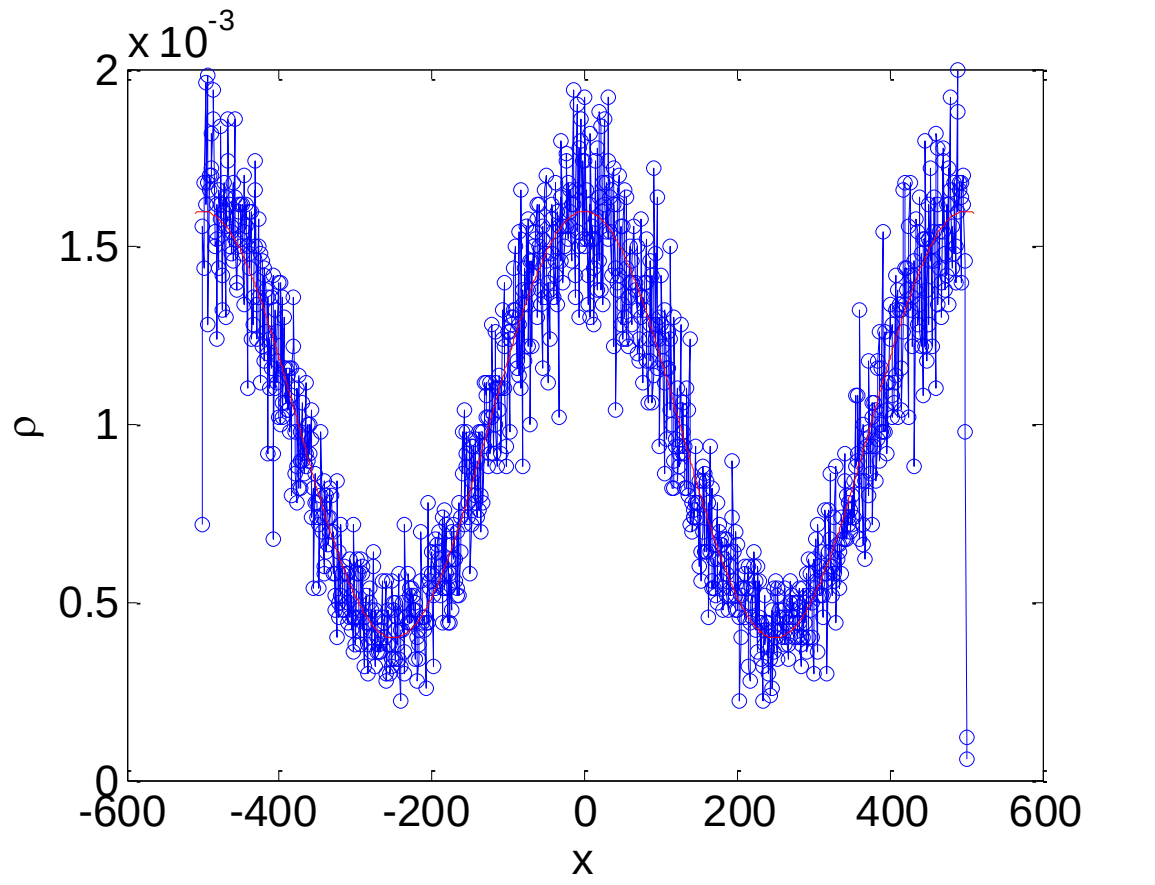

Figure 8: Arrival frequency (blue) and theoretical value (red) for $N_p = 50000$, $t = 500$ as a function of position (free particle, two sources, $D = 1$, $P_0^{(D)} = .1$).

**Multiple sources distanced by $a$**

A generalisation of the previous case to a scenario with an even number $N_s$ of sources distanced by the quantity $a$ in lattice units is formalised as $x_0 = \{ka\}$, with $k = \left(-\frac{N_s-1}{2}, \dots, \frac{N_s-1}{2}\right)$. The source probabilities are such that $\sum_{k=-\frac{N_s-1}{2}}^{\frac{N_s-1}{2}} P_0^{(k)} = 1$ and the source phase is $\epsilon^{(k)} \equiv 0$. In this case $N_s(N_s - 1)$ types of bosons arise, with $\delta^{(ij)} = |i-j|a$ and $\epsilon^{(ij)} = 0$. The theoretically expected pdf is thus

$$\rho(\boldsymbol{x};t) = \frac{1 + \sum_i \sum_{i \neq i} \sqrt{P_0^{(i)}P_0^{(j)}}\cos\left(\pi\delta^{(ij)}\frac{\boldsymbol{x} - \frac{a(i+j)}{2}}{t}\right)}{2t}. \qquad (64)$$

Figure 9 show the frequency of arrivals after $N_t = 500$ iterations with $a = 2$, $N_s = 3$, and $P_0^{(j)} = 1/N_s$, $\forall j$. The frequency clearly tends to the theoretical probability density (63). Figure Figure 10 shows an additional result, namely, the distribution of quantum momentum $v_Q$ as a function of the source momentum $v_0$. It is clearly this pattern that builds (64) via the chain rule (25).

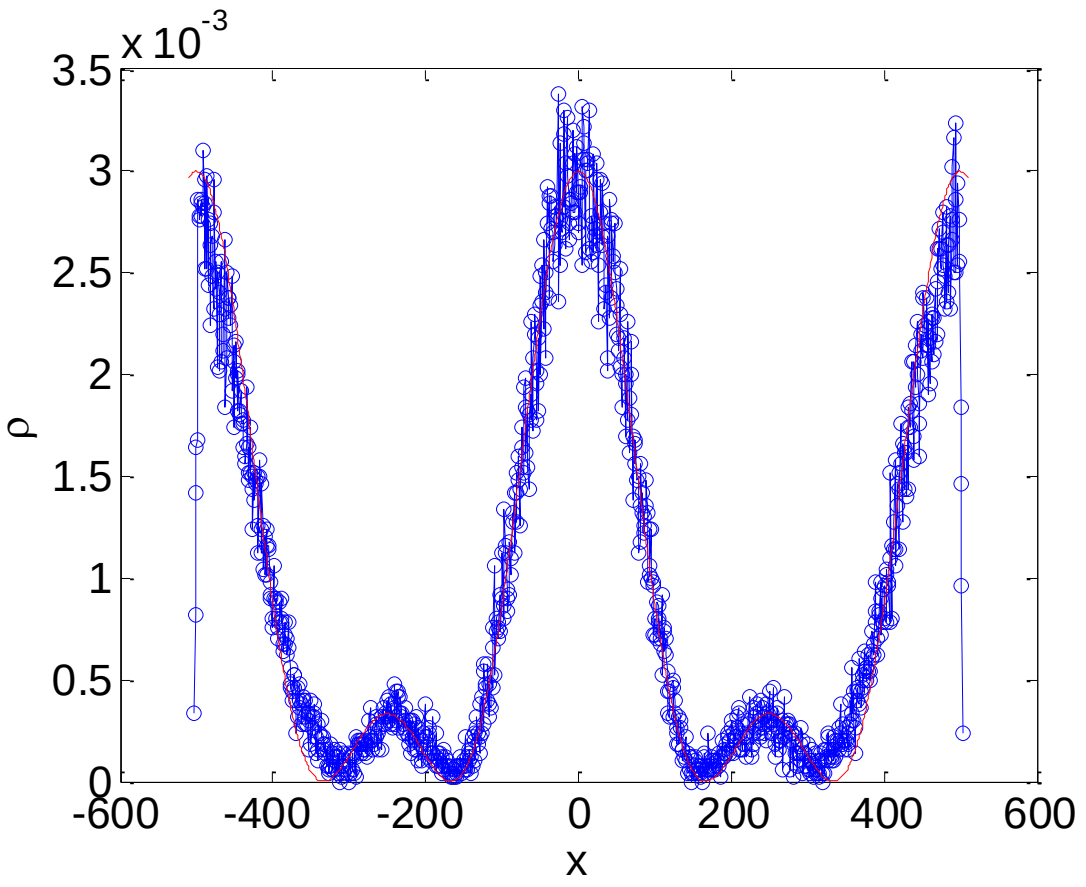

Figure 9: Arrival frequency (blue) and theoretical value (red) for $N_p = 50000$, $t = 500$ as a function of position (free particle, multiple sources, $a = 2$, $N_s = 3$).

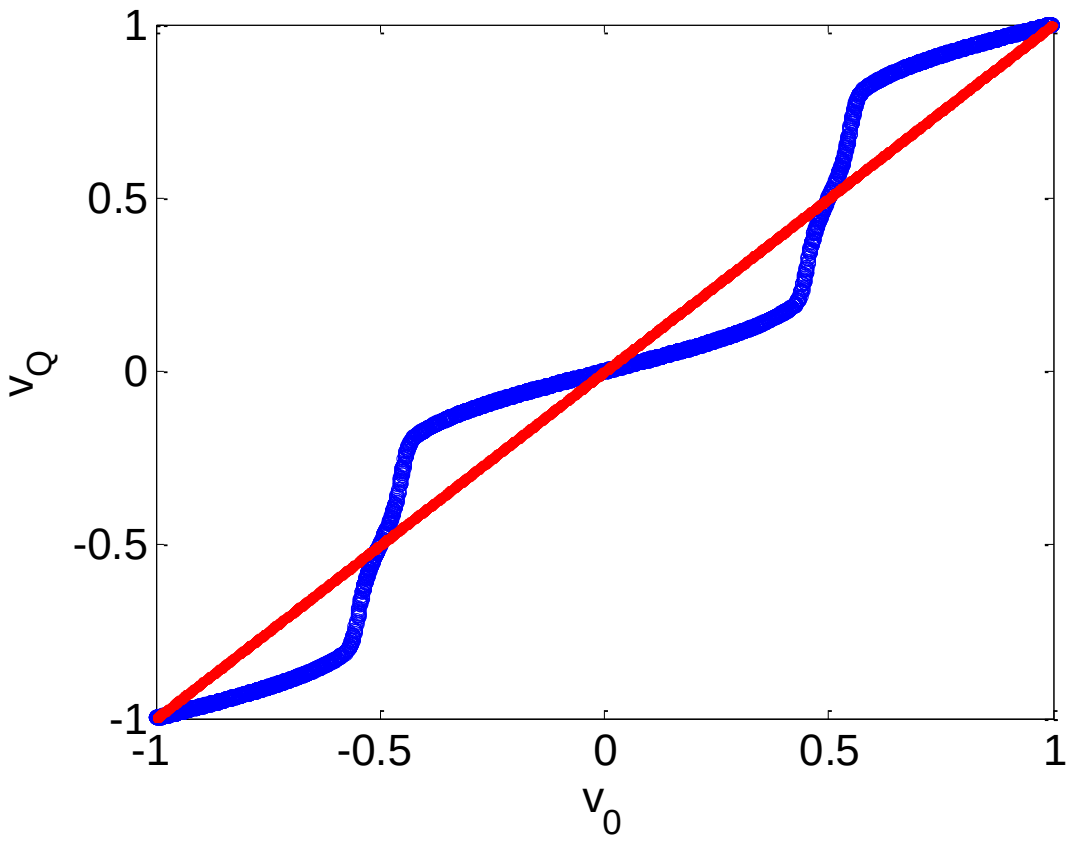

Figure 10: Quantum momentum distribution for $N_p = 50000$, $t = 500$ as a function of source momentum (free particle, multiple sources, $a = 2$, $N_s = 3$).

**Plane wave**

A scenario that resembles a "plane wave" state is obtained by placing $\ell + 1$ equiprobable sources distanced by one lattice node ($x_0 = \{k\}$, $k \in [-\ell/2, \ell/2]$, $P_0^{(k)} = 1/(\ell + 1)$), and providing these sources with a phase momentum $\epsilon^{(k)} = k v_\varphi$. As each plane wave is a stationary state for the free quantum particle, the theoretically expected pdf is

$$\rho(\boldsymbol{x}; t) \approx \frac{1}{\ell + 1}, \qquad x \in \left[-\ell/2 + v_\varphi t, \ell/2 + v_\varphi t\right]. \tag{65}$$

for sufficiently large $\ell$; however, for finite $\ell$ the theoretical value results from the application of the propagator (59) to a state $\psi_0 = e^{\iota \pi v_\varphi x}/\sqrt{\ell + 1}$ and is generally different from (65).

Figure 11 shows the frequency of arrival after $N_t = 500$ iterations with $\ell = 100$ and $v_\varphi = 0.1$.

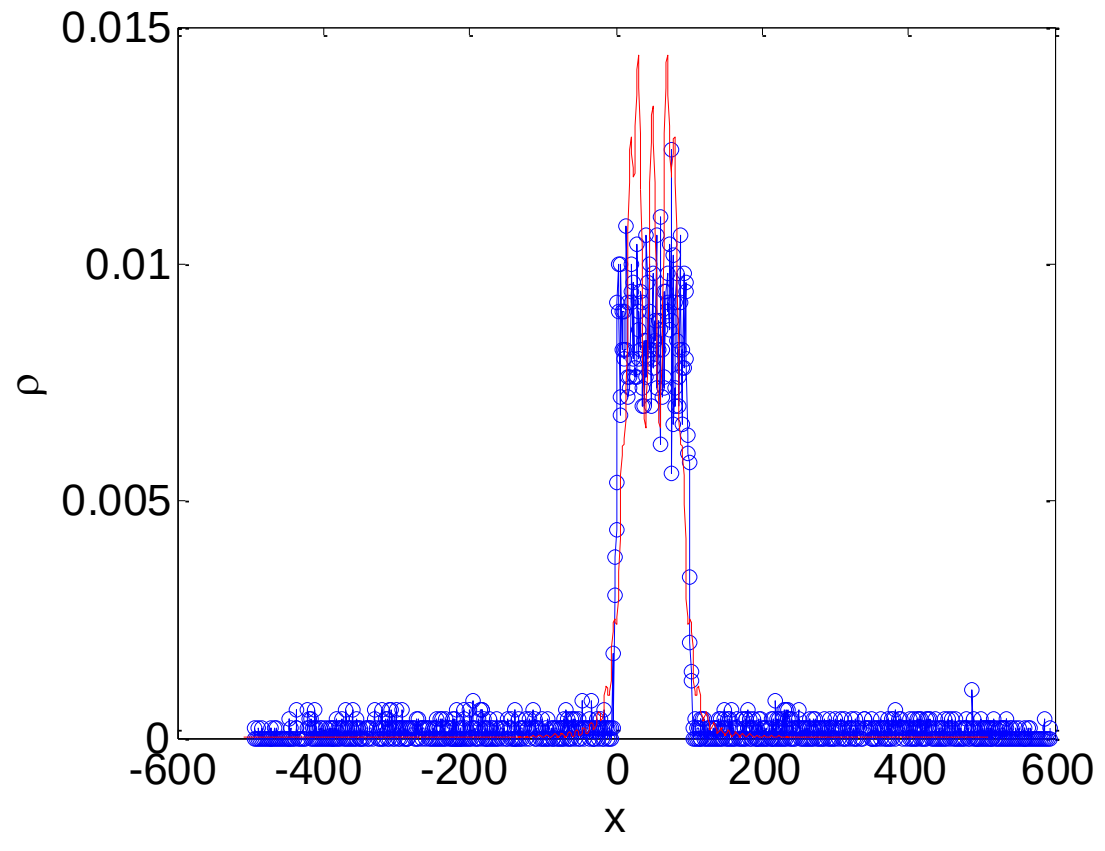

Figure 11: Arrival frequency (blue) and theoretical value (red) for $N_p = 5000$, $t = 500$ as a function of position (free particle, plane wave, $\ell = 100$, $v_\varphi = 0.1$).

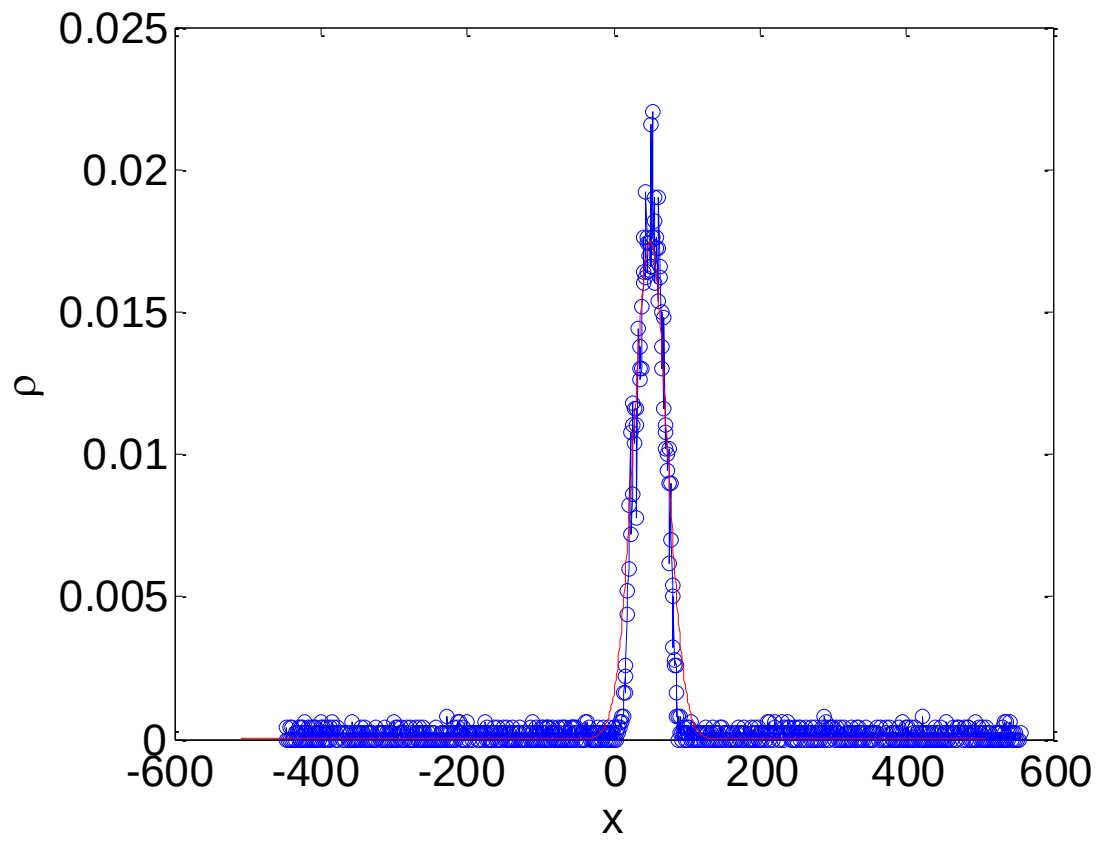

Figure 12: Arrival frequency (blue) and theoretical value (red) for $N_p = 50000$, $t = 500$ as a function of position (free particle, Gaussian wave, $a = 0$, $d = 5$, $v_\varphi = 0.1$).

### Gaussian wave

This scenario implies several sources with different probabilities and a phase momentum. Namely, $x_0 = \{k\}$, $k \in (-\infty, \infty)$, with $P_0^{(k)} = \left(\frac{1}{\pi d^2}\right)^{\frac{1}{2}} \exp\left(-\frac{(k-a)^2}{d^2}\right)$ and $\epsilon^{(k)} = kv_\varphi$. The theoretically expected pdf is obtained by applying the propagator (59) to an initial state $\psi_0(y) = (1/\pi d^2)^{1/4} \exp\left(-(y-a)^2/d^2 + \iota\pi v_\varphi (y-a)\right)$ and is calculated as

$$\rho(\boldsymbol{x}; t) = \left(\frac{1}{\pi d^2 \left(1 + \frac{t^2}{\pi^2 d^4}\right)}\right)^{\frac{1}{2}} \exp\left(-\frac{\left(x - a - v_\varphi t\right)^2}{d^2 \left(1 + \frac{t^2}{\pi^2 d^4}\right)}\right). \tag{66}$$

Figure 12 shows the frequency of arrival after $N_t = 500$ iterations with $a = 0$, $v_\varphi = 0.1$, and $d = 5$.

## A.2 Free faller

For all the scenarios in this section, $f(x) \equiv \phi$. Results are compared with quantum mechanics (theoretical values) by using the propagator

$$K^{(FF)}(x, t|y) = \frac{1}{\sqrt{2\iota t}} \exp\left(\frac{\iota\pi(x-y)^2}{2t} + \frac{\iota\pi(x+y)\phi t}{2} - \frac{\iota\pi\phi^2 t^3}{24}\right) \tag{67}$$

where the exponential argument is clearly the classical action (in lattice units) multiplied by the factor $\iota\pi$.

### Single source

In this case, $x_0 = \{x_0\}$, $P_0 = \{1\}$. The theoretically expected pdf is

$$\rho(\boldsymbol{x}; t) = \frac{1}{2t}, \qquad \boldsymbol{x} \in \left[x_0 - t + \frac{\phi t^2}{2}, x_0 + t + \frac{\phi t^2}{2}\right]. \tag{68}$$

Figure 13 shows the frequency of arrival after $N_t = 200$ iterations with $\phi = 2 \cdot 10^{-3}$ and $N_p = 50000$. Arrivals clearly tend toward distribution (68), with $x \in [-160,240]$.

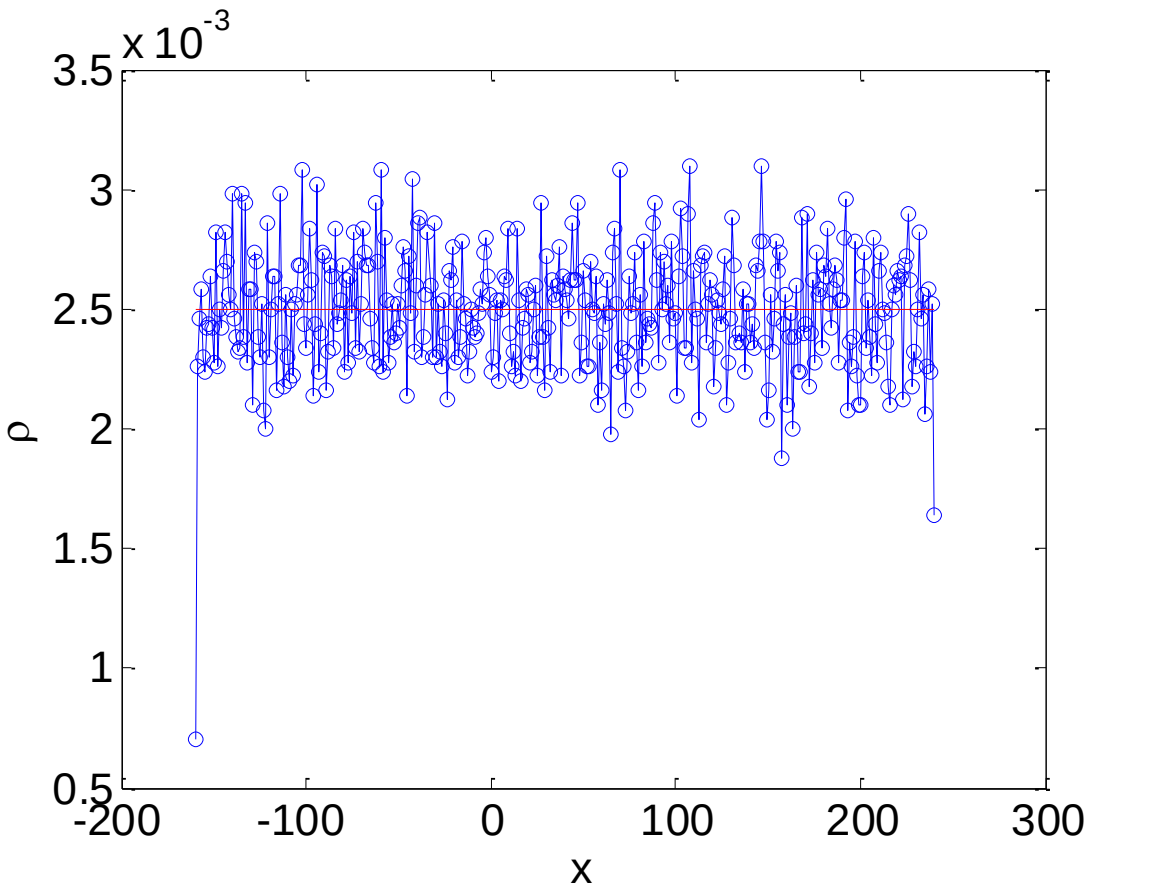

Figure 13: Arrival frequency (blue) and theoretical value (red) for $N_p = 50000$, $t = 200$ as a function of position (free faller, single source, $x_0 = 0$, $\phi = .002$).

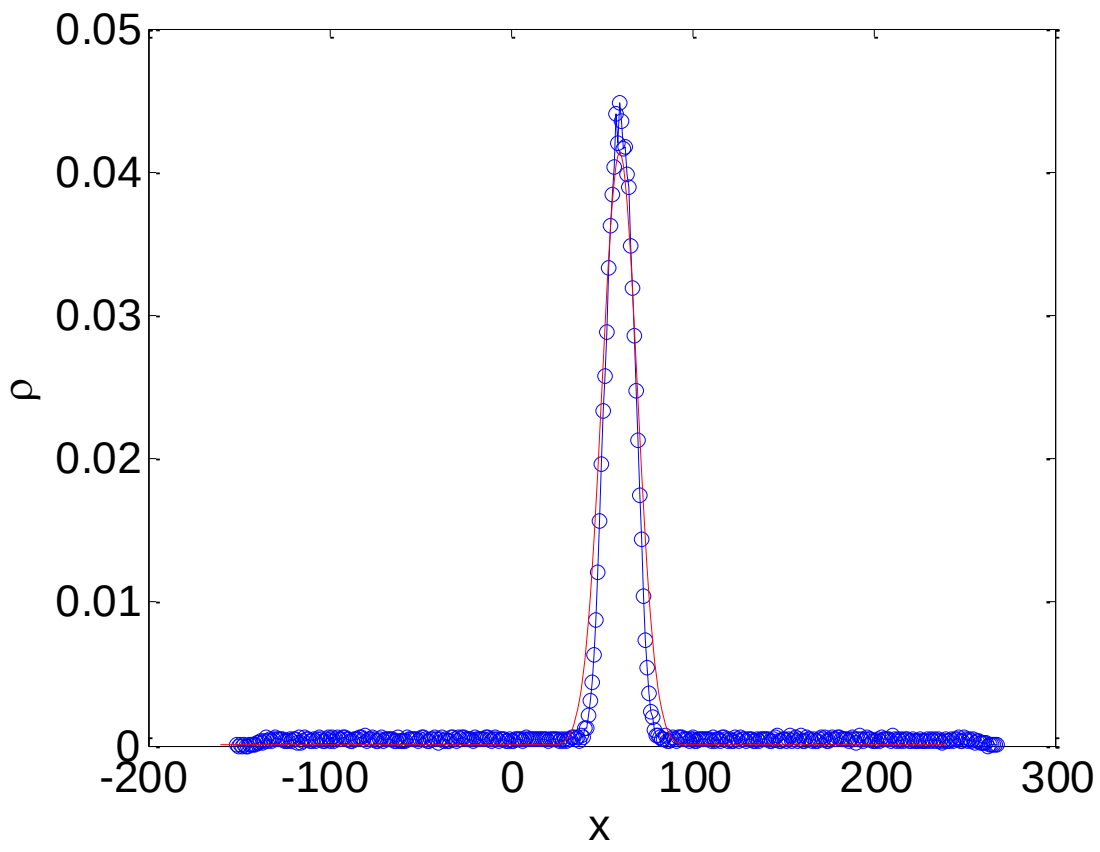

Figure 14: Arrival frequency (blue) and theoretical value (red) for $N_p = 50000$, $t = 200$ as a function of position (free faller, Gaussian wave, $\phi = .002$, $a = 0$, $d = 5$, $v_\varphi = 0.1$).

**Gaussian wave**

In this case the sources are prepared as in Sect. 0. The theoretically expected pdf is found by applying the propagator (67) to the initial state $\psi_0$ shown in that section and is

$$\rho(\boldsymbol{x};t) = \left(\frac{1}{\pi d^2\left(1+\frac{t^2}{\pi^2 d^4}\right)}\right)^{\frac{1}{2}} \exp\left(-\frac{\left(x-a-v_\varphi t-\frac{\phi t^2}{2}\right)^2}{d^2\left(1+\frac{t^2}{\pi^2 d^4}\right)}\right). \tag{69}$$

Figure 14 shows the frequency of arrival after $N_t = 200$ iterations with $\phi = 2\cdot 10^{-3}$, $a = 0$, $v_\varphi = 0.1$, $d = 5$, and $N_p = 50000$. The centre of the Gaussian wave, initially at $x = 0$, has moved to the left to the point $v_\varphi t + \phi t^2/2 = 60$.

## A.3 Harmonic oscillator

This scenario is described by distributing external bosons at each node of the lattice, all of them having a momentum $f(x) = -\Omega^2 x$. Results are compared with quantum mechanics (theoretical values) by using the propagator

$$K^{(HO)}(x,t|y) = \sqrt{\frac{\Omega}{2\iota\sin\omega t}}\exp\left(\frac{\iota\pi\Omega}{2\sin\Omega t}\left((x^2+y^2)\cos\Omega t - 2xx_0\right)\right) \tag{70}$$

where again (quadratic Lagrangian) the argument of the exponential is clearly the classical action multiplied by the factor $\iota\pi$.

**Single source**

In this case, $x_0 = \{x_0\}$, $P_0 = \{1\}$. The theoretically expected pdf is

$$\rho(\boldsymbol{x};t) = \frac{\Omega}{2\sin(\Omega t)}, \quad \boldsymbol{x} \in \left[x_0\cos(\Omega t) - \frac{1}{\Omega}\sin(\Omega t), x_0\cos(\Omega t) + \frac{1}{\Omega}\sin(\Omega t)\right]. \tag{71}$$

Figure 15 shows the frequency of arrival after $N_t = 200$ iterations with $\Omega = .005$ and $N_p = 50000$. Arrivals clearly tend toward distribution (71), with $x \in [-168,168]$.

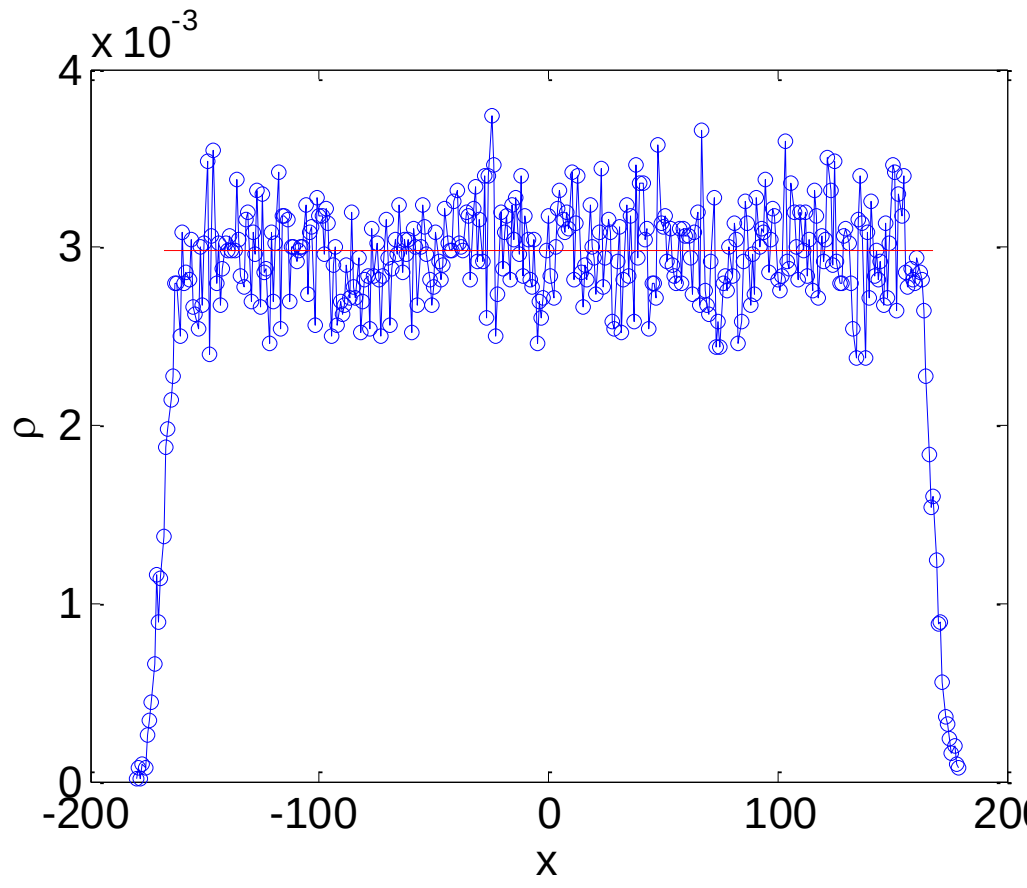

Figure 15: Arrival frequency (blue) and theoretical value (red) for $N_p = 50000$, $t = 200$ as a function of position (harmonic oscillator, single source, $\Omega = .005$,

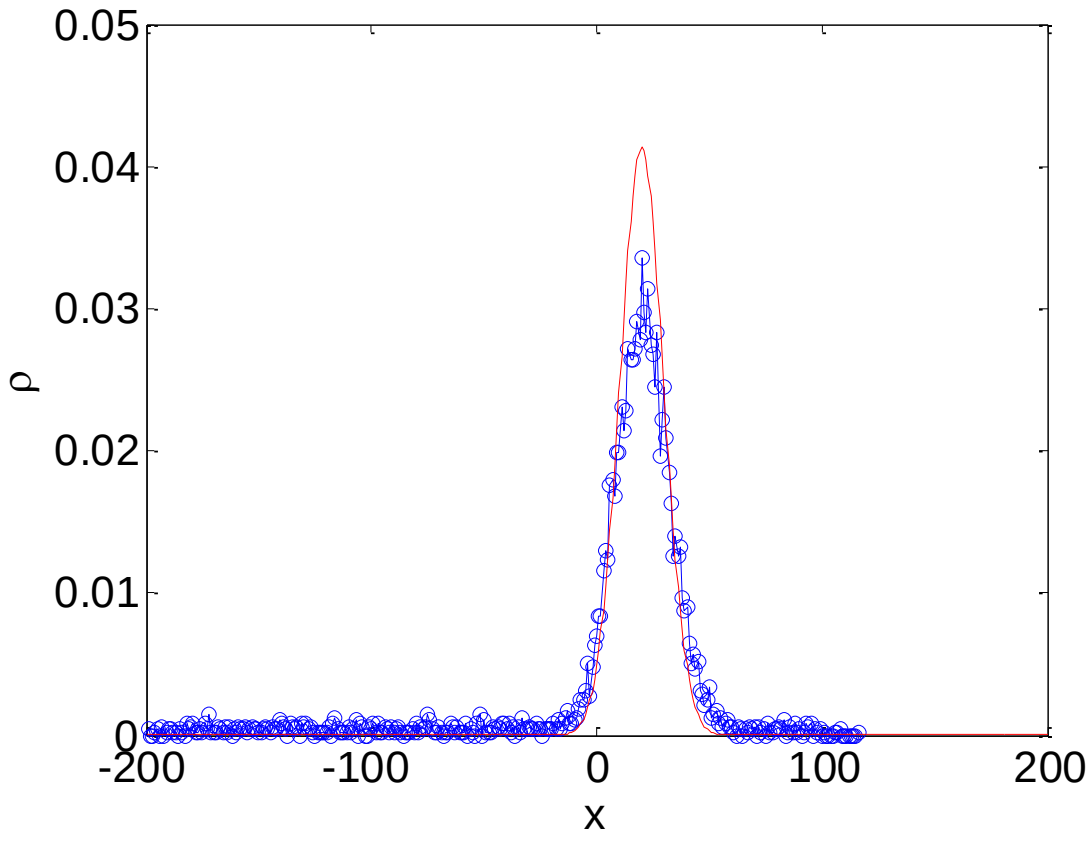

Figure 16: Arrival frequency (blue) and theoretical value (red) for $N_p = 5000$, $t = 200$ as a function of position (harmonic oscillator, Gaussian wave, $\Omega = 1\cdot 10^{-4}$, $a = 0$, $d = 5$, $v_\varphi = 0.1$).

$x_0 = 0$).

**Gaussian wave**

In this case the sources are prepared as in Sect. 0. The theoretically expected pdf is found by applying the propagator (70) to the initial state $\psi_0$ shown in that section and is

$$\rho(\boldsymbol{x};t) \approx \frac{1}{\sqrt{\pi}\,R} \exp\left( -\frac{\left(x - a - \frac{v_\varphi \sin \Omega t}{\Omega}\right)^2}{R^2} \right), \quad R \\ = \sqrt{d^2 \cos^2 \Omega t + \left(\frac{1}{\pi \Omega d}\right)^2 \sin^2 \Omega t}. \tag{72}$$

Figure 18 shows the frequency of arrival after $N_t = 200$ iterations with $\Omega = 1 \cdot 10^{-4}$, $a = 0$, $v_\varphi = 0.1$, $d = 5$, and $N_p = 5000$. The centre of the Gaussian wave, initially at $x = 0$, has moved to the left to the point $v_\varphi \sin \Omega t / \Omega = 20$.

**Stationary states**

In this case the source probability and the phase momentum are obtained from quantum mechanical initial states $\psi_0^{(n)}(y) = \Omega^{1/4} \frac{1}{\sqrt{2^n n!}} H_n\left(\sqrt{\pi\Omega} y\right) \exp\left(-\frac{\pi\Omega y^2}{2}\right)$, where $H_n$ are the Hermite polynomials. Since these are stationary states, the theoretically expected pdf is obtained as

$$\rho(\boldsymbol{x};t) \approx P_0^{(x)} = \Omega^{1/2} \frac{1}{2^n n!} H_n\left(\sqrt{\pi\Omega} x\right)^2 \exp(-\pi\Omega x^2). \tag{73}$$

Figure 17 shows the frequency of arrival after $N_t = 200$ iterations with $\Omega = 1 \cdot 10^{-4}$, $n = 1$, and $N_p = 5000$.

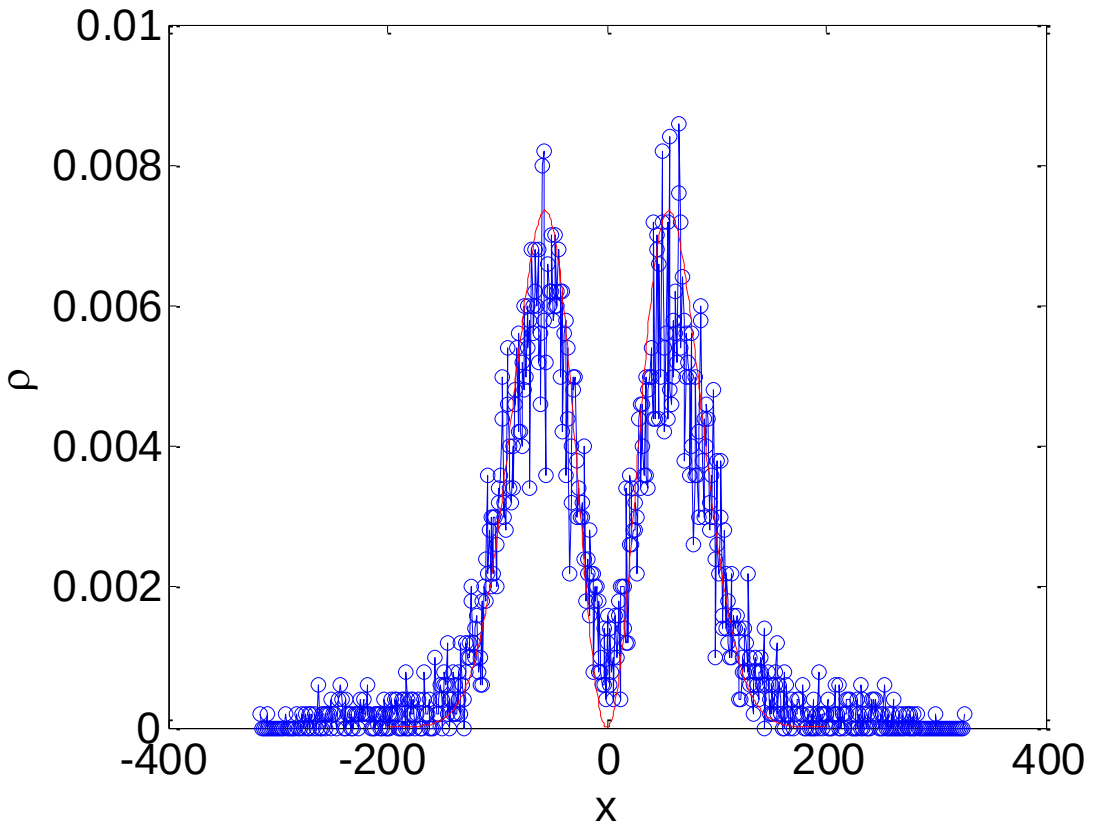

Figure 17: Arrival frequency (blue) and theoretical value (red) for $N_p = 5000$, $t = 200$ as a function of position (harmonic oscillator, stationary state, $\Omega = .0001$, $n = 1$).

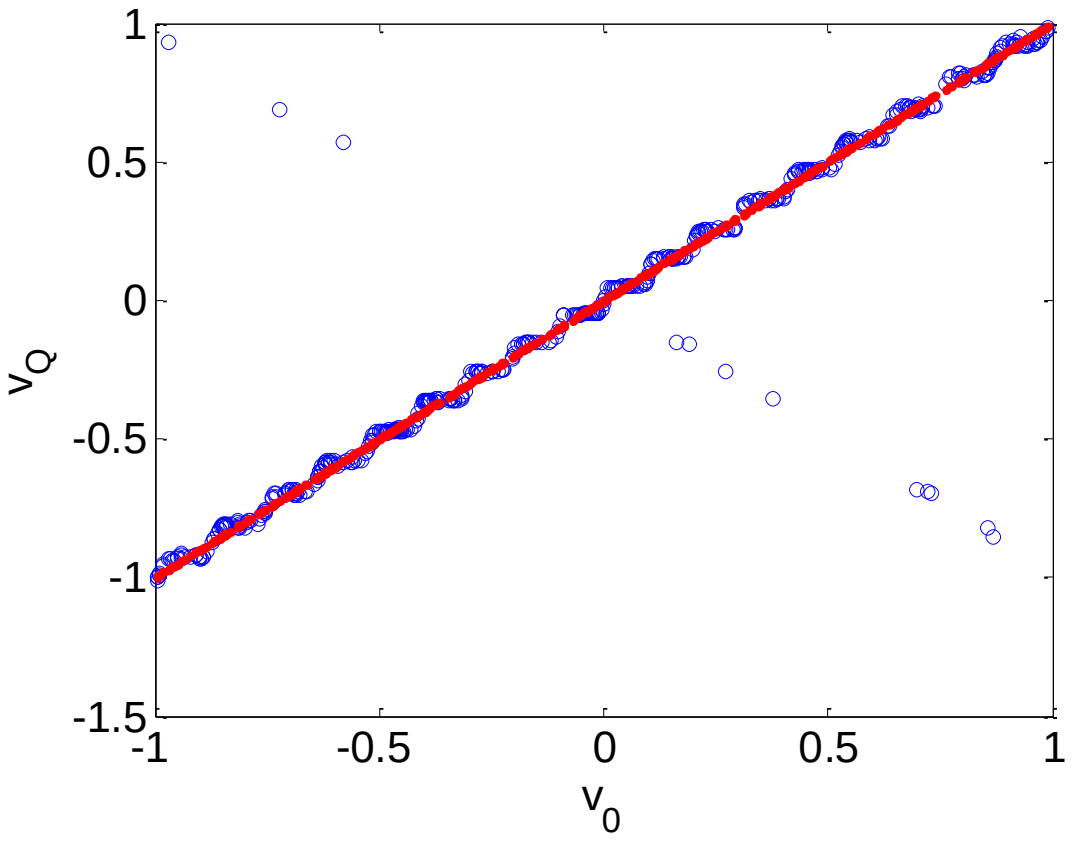

Figure 18: Quantum momentum distribution for $N_p = 500$, $t = 500$ as a function of source momentum (particle in a box, single source, $a = 10$, $x_0 = 0$).

## A.4 Particle in a box

This scenario is defined by the box size $a$, such that $x \in [-a, a]$, with $f = 0$. Each time a particle hits the box boundaries, it gains an external boson $v_F = -2v_Q$ (infinite potential outside the box).

Moreover, quantum forces arise even in the presence of a single source because of span re-initialization mechanism (13) that occurs at each boundary hit. Equation (46) is replaced by

$$\ell[n] = \frac{x[n] - (-1)^h x_0 \pm 2ha}{t[n]}, \tag{74}$$

where $h$ is the number of hits and the sign of the last term in the numerator depends on the direction of the first hit. A given node can be thus visited by particles having the same lifetime but different numbers of hits.

Results are compared with quantum mechanics (theoretical values) by using the propagator

$$K^{(BX)}(x,t|y) = \sum_{n=1}^{\infty} \exp\left(-\frac{iE_n t}{\hbar}\right) \psi^{(n)}(y)\psi^{(n)}(x), \qquad E_n = \frac{\hbar^2\pi^2 n^2}{8ma^2}. \tag{75}$$

where the stationary states are

$$\psi^{(n)}(x) = \sqrt{\frac{1}{a}} \cdot \begin{cases} \cos\frac{n\pi x}{2a}, & n = 1,3,5,\ldots \\ \sin\frac{n\pi x}{2a}, & n = 2,4,6,\ldots \end{cases} \tag{76}$$

Propagator (75) is equivalent to

$$K^{(BX)}(x,t|y) = \lim_{N_s\to\infty} \frac{1}{\sqrt{2\iota t}} \cdot \sum_{l=-N_s}^{N_s} (-1)^l \exp\left(\iota\pi \frac{\left(x - 2la - (-1)^l y\right)^2}{2t}\right), \tag{77}$$

that is, the propagator of a free particle with $2N_s + 1$ equally probable sources ("virtual sources"), equally separated by $2a \pm 2y$, where $N_s \approx t/(2a)$.

**Single source**
In this case, $x_0 = \{x_0\}$, $P_0 = \{1\}$. The theoretically expected pdf for the momentum is

$$\rho(\boldsymbol{x};t) = \frac{1}{2M+1} \sum_{m=0}^{M} \delta(x - \hat{x}^{(m)}). \tag{78}$$

where $\hat{x}^{(m)} = 2a/\pi \arcsin\left(\sin \pi\hat{v}^{(m)} t/(2a)\right)$ and $\hat{v}^{(m)} = \pm(2m + m_0)/(2a)$, with $m_0 = 0$ or 1 and $M = a - m_0$.

Figure 18 shows the distribution of the quantum momentum after $N_t = 500$ iterations with $a = 10$ and $N_p = 500$. Momentum tends to assume definite values $\pm 1/(2a)$, $\pm 3/(2a)$, ...., i.e., values $\hat{v}^{(m)}$ with $m_0 = 1$.

Remark: to see the correct interference pattern building in the spatial domain, one would require large $a$ and $t$, which makes large $N_s$ and consequently too long simulations to be afforded.

**Stationary state**
In this case, the source probability and phase momentum are obtained from quantum mechanical initial states (76). Since these are stationary states, the theoretically expected pdf is obtained as

$$\rho(\boldsymbol{x};t) \approx P_0^{(x)} = \frac{1}{a} \cdot \begin{cases} \cos^2\left(\frac{n\pi x}{2a}\right), & n = 1,3,5,\dots \\ \sin^2\left(\frac{n\pi x}{2a}\right), & n = 2,4,6,\dots \end{cases}, \tag{79}$$

that is, a function with peaks at $\hat{x}^{(m)} = \pm a/n(2m + m_0)$, with $m_0 = 0$ for $n$ odd and $m_0 = 1$ for $n$ even. Correspondingly, momentum distribution peaks at $\hat{v}^{(m)} = \pm n/(2a)$.

Figure 19-Figure 20 show the frequency of arrival after $N_t = 500$ iterations and for $N_p = 5000$, with $a = 10$ and $n = 3,\ 5$, respectively. Momentum clearly tends to the theoretically allowed values $\pm n/(2a)$.

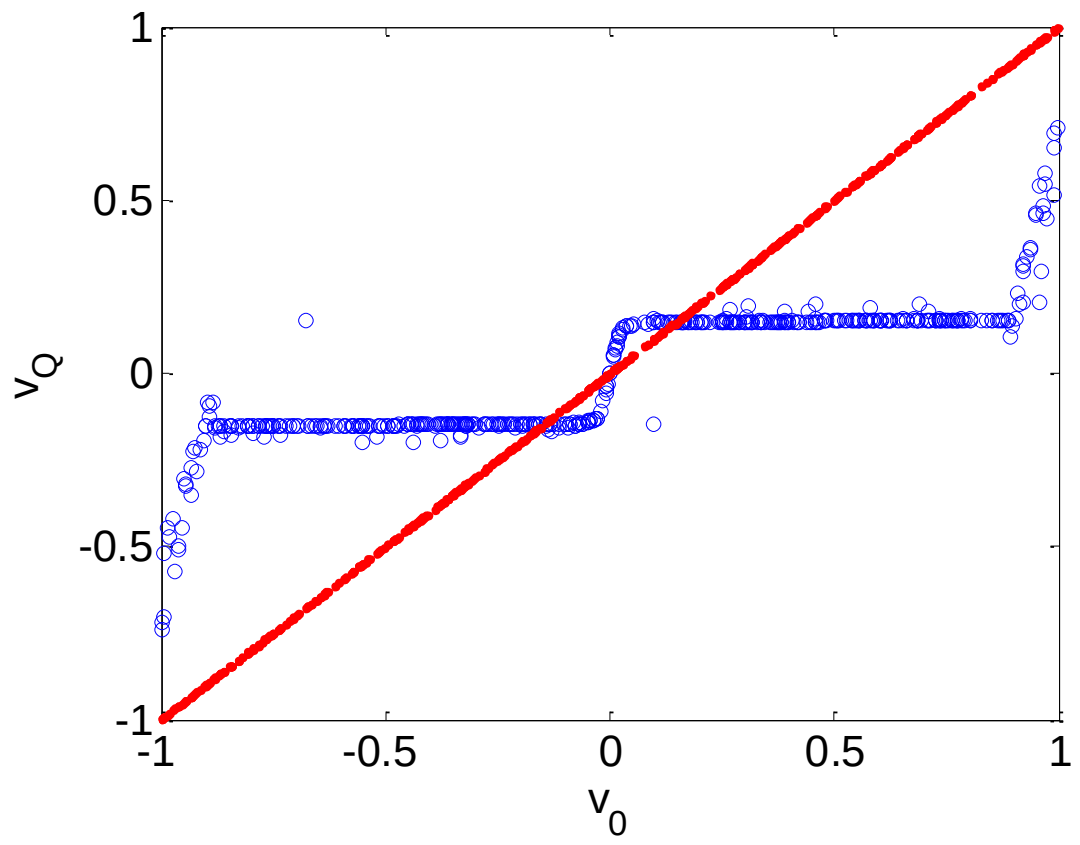

Figure 19: Quantum momentum distribution for $N_p = 500$, $t = 500$ as a function of source momentum (particle in a box, stationary state, $a = 10$, $n = 3$).

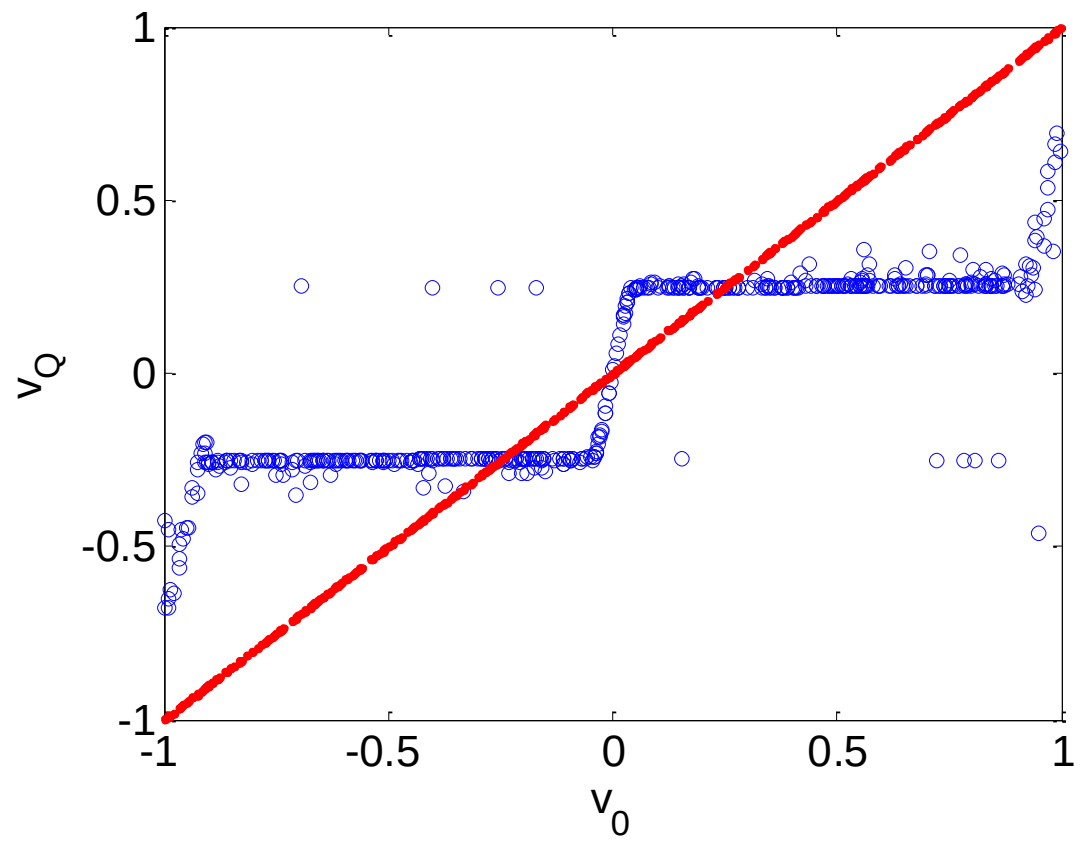

Figure 20: Quantum momentum distribution for $N_p = 500$, $t = 500$ as a function of source momentum (particle in a box, stationary state, $a = 10$, $n = 5$).

## A.5 Delta potential

This scenario is defined by the amplitude $\lambda$ of the Delta-type function (centred at $x = 0$) that describes the potential. This function is represented in the proposed model by setting $f(x) = \lambda/\ell$ for $x \in (-\ell, 0]$ and $f(x) = -\lambda/\ell$ for $x \in (-2\ell, -\ell]$, with $\ell$ an arbitrary scale (finite rectangular barrier).

Results are compared with quantum mechanics (theoretical values) by using the propagator

$$\begin{aligned} K^{(DP)}(x,t|y) &= \frac{1}{\sqrt{2\iota t}} \exp\left(\frac{\iota\pi(x-y)^2}{2t}\right) \\ &- \frac{\lambda\pi^2}{\sqrt{2\iota t}} \int_0^\infty du \exp\left(-u\pi^2\lambda - \frac{\pi(|x|+|y|+u)^2}{2\iota t}\right) = \\ &= K^{(FP)}(x,t|y) \\ &- \lambda\pi^2 \int_0^\infty du \exp(-u\pi^2\lambda) K^{(FP)}\big(x,t|{-x}/|x|(|y|+u)\big) \end{aligned} \tag{80}$$

that is clearly equivalent to that of a free particle with an infinity of "virtual sources" placed at$-\mathrm{sign}(x) \cdot (|y| + u)$, $u = 0, \dots$ Considering $y > 0$, when $x > 0$ these virtual sources are at $-y - u$, while for $x < 0$ they are at $y + u$. In both cases, they correspond to delayed bounces of the particle at the potential Delta.

In the proposed model, the virtual sources arise naturally as a consequence of the random motion around $x = 0$. In fact, particles can emerge from the potential Delta with several momentum values, depending on how much time they have spent in the finite rectangular barrier. With respect to the classical case where a particle bounces on the barrier if its energy is lower than the potential and its momentum is reverted ($v_0 \rightarrow -v_0$), here $v_0 \rightarrow v = (v_0 + f\tau^+ - f\tau^-)$, where $\tau^\pm$ is the time spent in the front, resp., rear side of the barrier. Assuming no quantum forces (see later), this momentum is kept in the absence of external forces and if, after $t$ iterations the particle has reached the node $x$, its overall trajectory is equivalent to that of a particle emitted at $x - vt$, that is, to a virtual source with at $u = -x_0 \pm (x + vt)$. As a consequence, the $u$'s are not integers but rational numbers as the $v_0$'s are.

Virtual sources can be lumped at their average value $-y - 1/\lambda\pi^2$ (with a $\pi$ phase) for $x > 0$ and $y + 1/\lambda\pi^2$ (with zero phase) for $x < 0$.

$$K^{(DP)}(x,t|y) \approx K^{(FP)}(x,t|y) - K^{(FP)}\left(x,t\middle| -\text{sign}(x)\cdot\left(|y| + \frac{1}{\lambda\pi^2}\right)\right) \tag{81}$$

**Single source**

In this case, $x_0 = \{x_0\}$, $P_0 = \{1\}$. By approximating the propagator (80) the theoretically expected pdf is approximated as

$$\rho(\boldsymbol{x};t) \approx \begin{cases} \frac{1}{2t}\cdot\frac{(x-x_0)^2}{\lambda^2\pi^2t^2 + (x-x_0)^2}, & x < 0 \\ \frac{1}{2t}\cdot\frac{(x+x_0)^2 + 2\lambda^2\pi^2t^2}{(x+x_0)^2 + \lambda^2\pi^2t^2}, & x > 0 \end{cases}. \tag{82}$$

From (82) the transmission ratio is retrieved (for $x_0 \ll t$) as

$$TRA := \int_{-t}^{0}\rho(\boldsymbol{x};t)dx \approx \frac{1}{2t}\left[x - \lambda\pi t\,\text{atan}\left(\frac{x}{\lambda\pi t}\right)\right]_{-t}^{0} = \frac{1}{2} - \frac{\lambda\pi}{2}\text{atan}\left(\frac{1}{\lambda\pi}\right), \tag{83}$$

and similarly the "reflection ratio" $REF = 1 - TRA$.

Figure 21 shows the reflection coefficient obtained for various values of $\lambda$ by summing the frequency of arrival at nodes $x < 0$ for $x_0 = 10$, $N_p = 5000$, $t = 200$, and $\ell = 15$. For simplicity the quantum forces have been disabled (the number of virtual sources and thus bosons to take into account would dramatically exceed the capabilities of a standard personal computer) as they do not contribute to the net transmission effect.

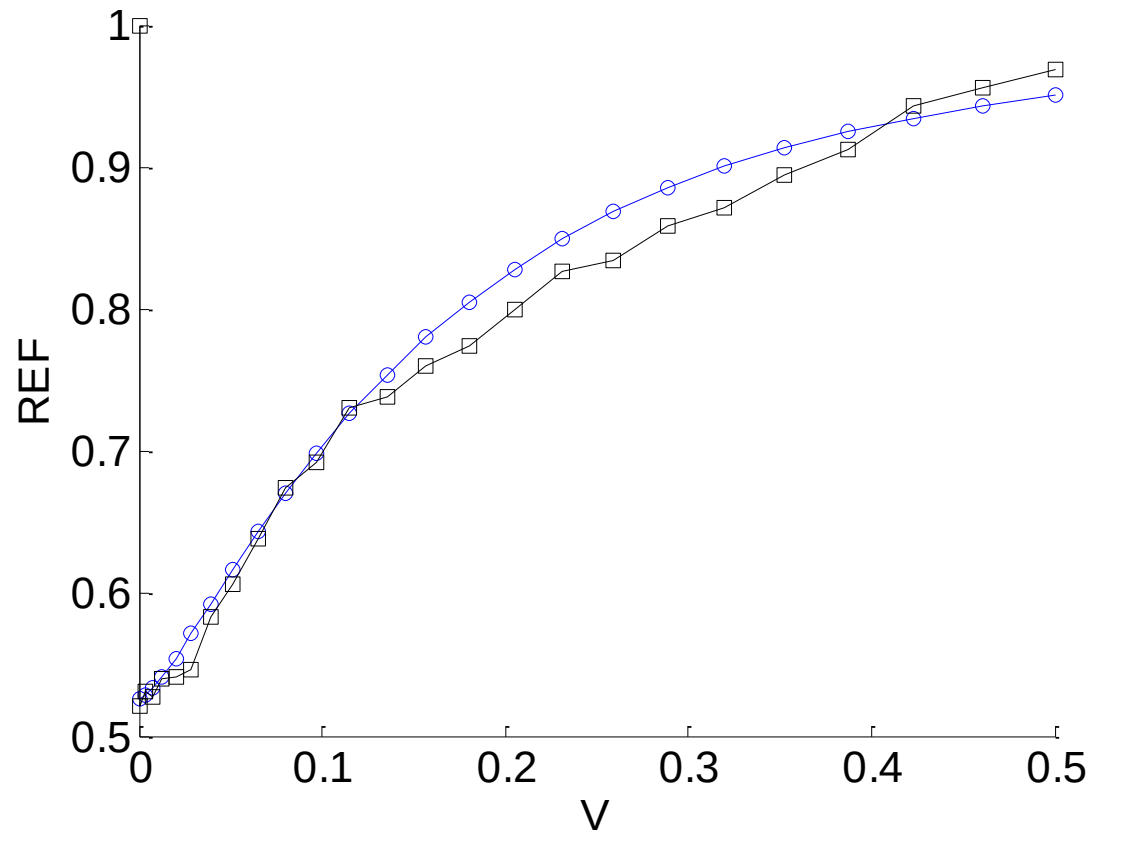

Figure 21: Reflection factor (black) and theoretical value (blue) for $N_p = 5000$, $t = 200$ as a function of

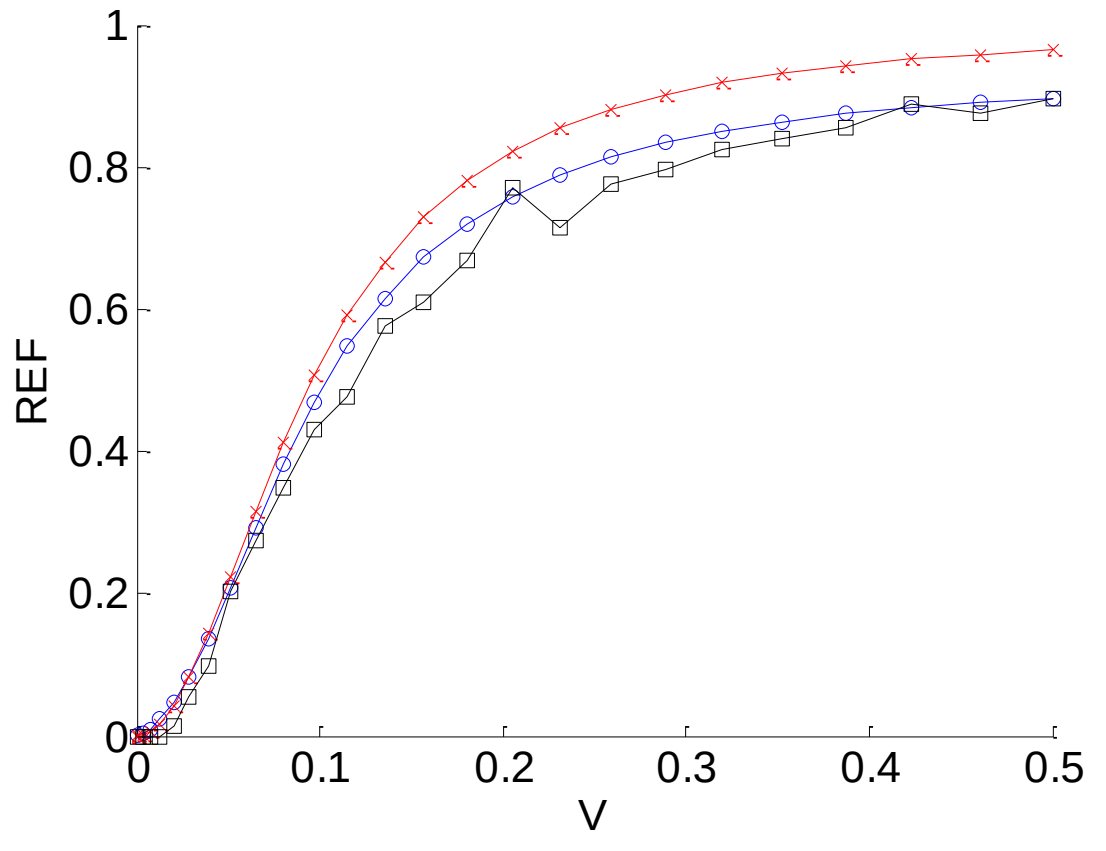

Figure 22: Reflection factor (black), theoretical value (blue) and approximation eq. (87) (red) for $N_p = 500$, $t = 3000$ as a function of potential $\lambda$ (Delta potential,

potential $\lambda$ (Delta potential, single source, $x_0 = 10$).    Gaussian wave, $a = 10$, $d = 10$, $v_\varphi = -0.3$).

**Gaussian wave**

In this case the sources are prepared as in Sect. 0. The theoretically expected pdf is found by applying the propagator (81) to the initial state $\psi_0$ shown in that section and is found as $\rho(\boldsymbol{x}; t) = |\psi(x,t)|^2$ where [Dodonov]

$$\psi(x,t) = \psi_0(x,t) \cdot \psi_1(x,t) \tag{84}$$

$$\psi_0(x,t) := \sqrt{\frac{d}{\sqrt{\pi}\mu(t)}} \exp\left[-\frac{\left(x - a - v_\varphi t\right)^2}{2\mu(t)} + \iota\pi v_\varphi x - \iota\pi v_\varphi^2 t/2\right] \tag{85}$$

$$\psi_1(x,t) := \left\{1 - \pi^2\lambda\sqrt{\frac{\pi\mu(t)}{2}} \operatorname{erfc}(D(x,t)) \exp\left(D(x,t)^2 + \frac{x + |x|}{\mu(t)}(-\iota\pi v_\varphi d^2 - a)\right)\right\} \tag{86}$$

where $\mu(t) \coloneqq d^2 + \iota t/\pi$ and $D(x,t) \coloneqq \left[\pi^2\lambda\mu(t) + (a + |x|) + \iota\pi v_\varphi d^2\right]/\sqrt{2\mu(t)}$.

For large times, an approximation of $\psi(x,t)$ allows to analytically evaluate the transmission coefficient that, if $d^2 v_\varphi^2 \gg 1$ further holds, reads

$$TRA \approx \frac{1}{1 + \left(\frac{\pi\lambda}{v_\varphi}\right)^2} \tag{87}$$

that is, the plane-wave transmission coefficient.

Figure 22 shows the reflection coefficient obtained for various values of $\lambda$ by summing the frequency of arrival at nodes $x < 0$ for $a = d = 10$, $v_\varphi = -0.3$, $N_p = 500$, $t = 3000$, and $\ell = 2$. For the same reason as in Sect. 0, the quantum forces due to virtual sources have been disabled as they do not contribute to the net transmission effect.